# Extended Fourier Analysis of Signals

## Dr. Sc. Comp. Vilnis Liepiņš

***Abstract***—*This summary of the doctoral thesis* **[9]** *provides a comprehensive formulation of the Extended Discrete Fourier Transform (EDFT), derived directly from the Fourier integral and its orthogonality properties. The method is obtained by solving weighted least-squares estimators in both continuous and discrete domains, yielding an adaptive frequency-domain representation that remains fully consistent with the classical Fourier framework. In the special case of uniformly sampled data on a uniform frequency grid of the same size, the EDFT reduces exactly to the classical Discrete Fourier Transform (DFT).*

*However, when the analysis grid exceeds the number of observed samples, EDFT circumvents conventional zero-padding by optimizing the transformation basis over the extended frequency set. This enables accurate spectral estimation from incomplete or nonuniformly sampled data. Consequently, the EDFT achieves enhanced frequency resolution in regions of strong spectral content while maintaining global resolution balance, thereby remaining consistent with the uncertainty principle. The inverse EDFT reconstructs the original signal and produces extrapolated or interpolated samples wherever spectral information is available.*

*The EDFT requires no explicit separation of deterministic and stochastic components and accurately captures broadband, transient, and sinusoidal features simultaneously. Simulation studies confirm its robustness under nonuniform sampling, multiple Nyquist zones, missing-data conditions, and signals with mixed spectra comprising both line and continuous components.*

*Although iterative computation of the EDFT entails higher numerical cost compared to the classical DFT, this limitation—significant in the 1990s—has been largely mitigated by modern computational resources, rendering the EDFT practical for contemporary signal analysis applications* **[11]**–**[58]**.

*In summary, the EDFT provides a powerful nonparametric framework for high-resolution spectral estimation under irregular sampling, incomplete data, and limited observation durations.*

## I. Introduction

The Fourier transform is a fundamental tool in signal analysis, providing a means to represent a real- or complex-valued time-domain function $x(t)$ (hereinafter referred to as the *signal*) in the frequency domain:

$$F(\omega) = \int_{-\infty}^{\infty} x(t) e^{-i\omega t} dt, \tag{1}$$

$$x(t) = \frac{1}{2\pi} \int_{-\infty}^{\infty} F(\omega) e^{i\omega t} d\omega. \tag{2}$$

The orthogonality property of the Fourier transform forms the basis of frequency-selective signal analysis:

$$\int_{-\infty}^{\infty} e^{-i\omega_0 t} e^{i\omega t} dt = 2\pi\delta(\omega - \omega_0), \tag{3}$$

where $\omega$ and $\omega_0$ denote cyclic frequencies, $i$ is the imaginary unit ($i^2 = -1$), and $\delta(\omega - \omega_0)$ is the Dirac delta function.

In practice, direct computation of (1) requires complete knowledge of the signal over the entire time axis and integration over an infinite interval, which is infeasible in real-world scenarios. Consequently, the observation period and integration limits are restricted to a finite interval Θ, and the signal is assumed known only within $-\Theta/2 \leq t \leq \Theta/2$.

The same constraint applies to discrete representations of the signal, such as nonuniformly sampled $x(t_k)$ or uniformly sampled $x(kT)$, where $k = -\infty, \ldots, -1, 0, 1, \ldots, +\infty$. In practice, only a finite-length sequence with $k = 0, 1, \ldots, K-1$, is available for analysis. Here, $K$ denotes the number of samples, $T$ the sampling period, and the total observation time is $\Theta = t_{K-1} - t_0 \approx KT$. To avoid aliasing and satisfy the Nyquist criterion, uniform sampling of a continuous-time signal must be performed with a period $T \leq \pi/\Omega$, where Ω is the highest angular frequency component of $x(t)$. Although nonuniform sampling does not impose such a strict constraint on the mean sampling period $T_s = \Theta/K$, it is assumed in subsequent analysis that both sampled sequences, $x(t_k)$ and $x(kT)$, originate from a band-limited signal $x(t)$ with bandwidth Ω.

In the following sections, the fundamental expressions for both the classical and the *extended Fourier analysis* of the continuous-time signal $x(t)$ and its sampled versions $x(t_k)$ and $x(kT)$ are derived and discussed.

# II. Problem Formulation

"*The formulation of a problem is often more essential than its solution which may be merely a matter of mathematical or experimental skill. To raise new questions, new possibilities, to regard old problems from a new angle requires creative imagination and marks real advances in science.*"
—Albert Einstein and Léopold Infeld, *The Evolution of Physics* (1938)

## A. Fundamental Expressions of Classical Fourier Analysis

Classical Fourier analysis addresses finite-time signals by employing truncated versions of the continuous Fourier transform. The corresponding finite-time transforms are defined as:

$$F_\Theta(\omega) = \int_{-\Theta/2}^{\Theta/2} x(t)e^{-i\omega t}dt, \tag{4.1}$$

$$F_\Theta(\omega) = \sum_{k=0}^{K-1} x(t_k)e^{-i\omega t_k}, \tag{4.2}$$

$$F_\Theta(\omega) = \sum_{k=0}^{K-1} x(kT)e^{-i\omega kT}. \tag{4.3}$$

The inverse transform for a band-limited signal is expressed as:

$$x_\Theta(t) = \frac{1}{2\pi}\int_{-\Omega}^{\Omega} F_\Theta(\omega)e^{i\omega t}d\omega. \tag{5}$$

Equations (4.2) and (4.3) represent the Discrete-Time Fourier Transforms (DTFT) of nonuniformly and uniformly sampled signals, respectively. The reconstructed signal $x_\Theta(t)$ outside the observation interval Θ decays rapidly, approaching values close to zero.

The amplitude spectrum of the signal is defined as the normalized Fourier transform:

$$S_\Theta(\omega) = \frac{1}{\Theta} F_\Theta(\omega). \tag{6}$$

In classical Fourier analysis, the achievable frequency resolution (referred to as *normal resolution*) is inversely proportional to the observation interval Θ; hence, longer observation periods yield higher spectral resolution.

Equation (4.1) can be derived directly from (1) by truncating the infinite integration limits, while (4.2) and (4.3) result from replacing infinite summations with finite ones. This formulation implicitly assumes that the signal $x(t)$ is zero outside the interval Θ. Consequently, the Fourier transform expressions in (4) are rigorously valid only for time-limited signals.

However, a signal that is strictly band-limited in Ω cannot simultaneously be time-limited; by definition, it must possess nonzero values outside the finite observation window Θ. Therefore, classical Fourier analysis provides only an approximation of (1), with accuracy improving as $\Theta \to \infty$. For any finite Θ, alternative transformation bases may exist that yield a more accurate estimation of the true continuous Fourier transform.

## B. Extended Fourier Analysis Formulation

The concept of *extended Fourier analysis* is to determine a transform basis applicable to band-limited signals observed over a finite interval Θ, yielding results that best approximate the ideal Fourier transform (1) in the least-squares ($L^2$) sense. The generalized transform expressions are defined as

$$F_\alpha(\omega) = \int_{-\Theta/2}^{\Theta/2} x(t)\alpha(\omega, t)dt, \tag{7.1}$$

$$F_\alpha(\omega) = \sum_{k=0}^{K-1} x(t_k)\alpha(\omega, t_k), \tag{7.2}$$

$$F_\alpha(\omega) = \sum_{k=0}^{K-1} x(kT)\alpha(\omega, kT), \tag{7.3}$$

with the inverse transform

$$x_\alpha(t) = \frac{1}{2\pi} \int_{-\Omega}^{\Omega} F_\alpha(\omega) e^{i\omega t}\, d\omega. \tag{8}$$

In general, the bases $\alpha(\omega, t)$, $\alpha(\omega, t_k)$, and $\alpha(\omega, kT)$ differ from the classical exponential form. The inverse transform (8) retains the exponential kernel and satisfies the Parseval–Plancherel equality:

$$\int_{-\infty}^{\infty} |x_\alpha(t)|^2 dt = \frac{1}{2\pi} \int_{-\Omega}^{\Omega} |F_\alpha(\omega)|^2\, d\omega. \tag{9}$$

To ensure that $F_\alpha(\omega)$ approximates $F(\omega)$ as closely as possible, the mean-square error is minimized:

$$|F(\omega) - F_\alpha(\omega)|^2 \to min. \tag{10}$$

Since the target $F(\omega)$ cannot be computed directly for a band-limited signal $x(t)$, it is substituted with a known model signal

$$x(\omega_0, t) = S(\omega_0) e^{i\omega_0 t}, \quad -\infty < t < \infty, \tag{11}$$

for frequencies in the range $|\omega_0| \le \Omega$. Its Fourier transform is expressed by the Dirac delta function defined by integral (3):

$$\int_{-\infty}^{\infty} x(\omega_0, t)e^{-i\omega t}dt = 2\pi S(\omega_0)\delta(\omega - \omega_0). \quad (12)$$

By applying (11) to (7) and inserting (12) into (10), we obtain the integral least-squares estimators for continuous, nonuniformly sampled, and uniformly sampled signals,

$$\Delta = \int_{-\Omega}^{\Omega} \left| 2\pi S(\omega_0)\delta(\omega - \omega_0) - \int_{-\Theta/2}^{\Theta/2} S(\omega_0)e^{i\omega_0 t}\alpha(\omega, t)dt \right|^2 d\omega_0, \quad (13.1)$$

$$\Delta = \int_{-\Omega}^{\Omega} \left| 2\pi S(\omega_0)\delta(\omega - \omega_0) - \sum_{k=0}^{K-1} S(\omega_0)e^{i\omega_0 t_k}\alpha(\omega, t_k) \right|^2 d\omega_0, \quad (13.2)$$

$$\Delta = \int_{-\Omega}^{\Omega} \left| 2\pi S(\omega_0)\delta(\omega - \omega_0) - \sum_{k=0}^{K-1} S(\omega_0)e^{i\omega_0 kT}\alpha(\omega, kT) \right|^2 d\omega_0. \quad (13.3)$$

The solutions of (13) yield the extended Fourier transform bases $\alpha(\omega, t)$, $\alpha(\omega, t_k)$, and $\alpha(\omega, kT)$ for each case.

To evaluate how closely the amplitude $S(\omega_0)$ of the model matches the signal's spectrum $S_\alpha(\omega)$, it is estimated as the ratio of the extended Fourier transforms (7) to the transform of a unit-amplitude exponential in the same basis:

$$S_\alpha(\omega) = \frac{\int_{-\Theta/2}^{\Theta/2} x(t)\alpha(\omega, t)dt}{\int_{-\Theta/2}^{\Theta/2} e^{i\omega t}\alpha(\omega, t)dt}, \quad (14.1)$$

$$S_\alpha(\omega) = \frac{\sum_{k=0}^{K-1} x(t_k)\alpha(\omega, t_k)}{\sum_{k=0}^{K-1} e^{i\omega t_k}\alpha(\omega, t_k)}, \quad (14.2)$$

$$S_\alpha(\omega) = \frac{\sum_{k=0}^{K-1} x(kT)\alpha(\omega, kT)}{\sum_{k=0}^{K-1} e^{i\omega kT}\alpha(\omega, kT)}. \quad (14.3)$$

For the exponential basis $\alpha(\omega, t) = e^{-i\omega t}$, the denominator in (14.1) equals Θ, recovering the classic case (6). In general, the denominator in formulas (14) is inversely proportional to the frequency resolution of the extended Fourier transform.

Assuming a *unit-amplitude model* $S(\omega_0) = 1$ for $|\omega_0| \le \Omega$ and zeros otherwise, the estimators reduce to:

$$\Delta = \int_{-\Omega}^{\Omega} \left| 2\pi\delta(\omega - \omega_0) - \int_{-\Theta/2}^{\Theta/2} e^{i\omega_0 t}\alpha(\omega, t)dt \right|^2 d\omega_0, \quad (15.1)$$

$$\Delta = \int_{-\Omega}^{\Omega} \left| 2\pi\delta(\omega - \omega_0) - \sum_{k=0}^{K-1} e^{i\omega_0 t_k}\alpha(\omega, t_k) \right|^2 d\omega_0, \quad (15.2)$$

$$\Delta = \int_{-\Omega}^{\Omega} \left| 2\pi\delta(\omega - \omega_0) - \sum_{k=0}^{K-1} e^{i\omega_0 kT}\alpha(\omega, kT) \right|^2 d\omega_0, \quad (15.3)$$

where the target distribution is expressed by the delta function and the synthesis estimators are determined by integral (15.1) or by discrete sums (15.2), (15.3).

The solutions (15) establish a formal and mathematically rigorous relationship between the classical and extended Fourier transforms. They provide the foundation for further generalization with (13), algorithmic optimization, and practical implementation in modern signal analysis frameworks.

# III. Problem Solution

In this section, the continuous and discrete-time synthesis estimators (13) and (15) are solved using a weak-formulation framework, and the resulting linear equation systems are analyzed to identify solutions suitable for practical implementation.

We adopt the following matrix conventions: superscripts (−1), T, *, and H denote matrix inverse, transpose, complex conjugate, and Hermitian transpose, respectively. The operator "**./**" denotes element-wise division, while *diag*(**X**) extracts the main diagonal of a square matrix **X** (or forms a diagonal matrix from a vector).

As established in the Appendices, directly minimizing an $L^2$ error functional (15) containing a Dirac delta distribution is not strictly defined because the square $|\delta|^2$ diverges. To resolve this, we interpret the relation between the target distribution and the integral or discrete synthesis estimators $I(\omega_0)$ in the weak sense. This is achieved by projecting the synthesis equations onto an appropriate set of test functions $\varphi(\omega_0)$, forcing the approximation projection to match the projection of the target distribution over the optimization band $\omega_0 \in [-\Omega, \Omega]$:

$$\int_{-\Omega}^{\Omega} I(\omega_0)\varphi(\omega_0)d\omega_0 = \int_{-\Omega}^{\Omega} 2\pi\delta(\omega-\omega_0)\varphi(\omega_0)d\omega_0,$$

where $I(\omega_0)$ taken from (15). Depending on the signal type—$x(t)$, $x(t_k)$ and $x(kT)$—the weak formulation is constructed by selecting the test function space to match the corresponding continuous or discrete Fourier basis functions indexed by the time parameters $\tau$, $t_l$, or $lT$:

$$\varphi(\omega_0) = e^{-i\omega_0\tau}, \quad -\Theta/2 \le \tau \le \Theta/2 \tag{16.1}$$

$$\varphi_l(\omega_0) = e^{-i\omega_0 t_l}, \quad l = 0, 1, \ldots, K-1 \tag{16.2}$$

$$\varphi_l(\omega_0) = e^{-i\omega_0 lT}, \quad l = 0, 1, \ldots, K-1 \tag{16.3}$$

Evaluating these weak-formulation identities leads directly to the following systems of linear integral and algebraic equations:

$$\int_{-\Theta/2}^{\Theta/2} \frac{\sin(\Omega(t-\tau))}{\pi(t-\tau)}\alpha(\omega,t)dt = e^{-i\omega\tau}, \tag{17.1}$$

$$\sum_{k=0}^{K-1} \frac{\sin(\Omega(t_k-t_l))}{\pi(t_k-t_l)}\alpha(\omega,t_k) = e^{-i\omega t_l}, \tag{17.2}$$

$$\sum_{k=0}^{K-1} \frac{\sin(\Omega(k-l)T)}{\pi(k-l)T}\alpha(\omega,kT) = e^{-i\omega lT}. \tag{17.3}$$

Detailed derivations of (17.1) – (17.3) using projection approach are provided in **APPENDIX A** and **APPENDIX B**.

For discreate-time signals $x(t_k)$ and $x(kT)$ with an arbitrary signal model $S(\omega_0)$, a weighted weak formulation is applied. By choosing a weighted set of complex exponentials matching the sampling grid, $\varphi_l(\omega_0) = S^*(\omega_0)e^{-i\omega_0 t_l}$ (or $\varphi_l(\omega_0) = S^*(\omega_0)e^{-i\omega_0 lT}$), as the test functions, we obtain the weak-form solutions. This yields the following systems of linear equations for $l = 0, 1, ..., K-1$:

$$\sum_{k=0}^{K-1}\left(\frac{1}{2\pi}\int_{-\Omega}^{\Omega}|S(\omega_0)|^2 e^{i\omega_0(t_k-t_l)}d\omega_0\right)\alpha(\omega,t_k) = |S(\omega)|^2 e^{-i\omega t_l}, \tag{18.1}$$

$$\sum_{k=0}^{K-1}\left(\frac{1}{2\pi}\int_{-\Omega}^{\Omega}|S(\omega_0)|^2 e^{i\omega_0(k-l)T}d\omega_0\right)\alpha(\omega,kT) = |S(\omega)|^2 e^{-i\omega lT}. \tag{18.2}$$

Here, $|S(\omega)|^2$ denotes the signal model power at $\omega_0 = \omega$. Detailed derivations for these weighted projections are provided in APPENDIX C.

## A. Extended Fourier Transform of Continuous-Time Signals

The solution of (17.1) employs a specific system of orthogonal functions—prolate spheroidal wave functions (PSWFs) $\psi_k(t)$, $k = 0, 1, 2, \ldots$—which form the basis of the extended Fourier transform [1]. The basis $\alpha(\omega, t)$ is expressed as a series expansion:

$$\alpha(\omega,t) = \sum_{k=0}^{\infty}\frac{B_k(\omega)}{\lambda_k}\psi_k(t). \tag{19}$$

A detailed step-by-step derivation of (19) is given in [5]. Accordingly, the extended Fourier transform of a continuous-time signal $x(t)$ is defined as:

$$F_\alpha(\omega) = \sum_{k=0}^{\infty} B_k(\omega)a_k\,, \quad |\omega| \le \Omega, \tag{20.1}$$

$$x_\alpha(t) = \sum_{k=0}^{\infty}\psi_k(t)a_k\,, \quad -\infty < t < \infty, \tag{20.2}$$

$$S_\alpha(\omega) = \frac{\sum_{k=0}^{\infty}B_k(\omega)a_k}{\sum_{k=0}^{\infty}|B_k(\omega)|^2}. \tag{20.3}$$

The coefficients and parameters are given by

$$a_k = \frac{1}{\lambda_k}\int_{-\Theta/2}^{\Theta/2}x(\tau)\psi_k(\tau)d\tau,\ \lambda_k = \int_{-\Theta/2}^{\Theta/2}\psi_k^2(t)dt,\ B_k(\omega) = \sqrt{\frac{\pi\Theta}{\lambda_k\Omega}}\,\psi_k(\omega\frac{\Theta}{2\Omega})(-i)^k\,.$$

The Parseval–Plancherel equality holds:

$$\int_{-\infty}^{\infty}|x_\alpha(t)|^2dt = \frac{1}{2\pi}\int_{-\Omega}^{\Omega}|F_\alpha(\omega)|^2\,d\omega = \sum_{k=0}^{\infty}|a_k|^2. \tag{21}$$

In practice, computation of (20.1) requires the evaluation of infinite series, which is infeasible for real-world applications. However, theoretical analysis shows that the denominator in the amplitude spectrum formula (20.3), $\sum_{k=0}^{K}|B_k(\omega)|^2$, tends to infinite as $K \to \infty$. Consequently, the extended Fourier transform exhibits *super-resolution*, i.e., the ability to resolve closely spaced sinusoids or complex exponentials whose frequencies differ by arbitrarily small finite values.

For numerical applications, the series in (19) must be truncated to a finite number $K$ of terms:

$$\alpha_K(\omega,t) = \sum_{k=0}^{K}\frac{B_k(\omega)}{\lambda_k}\psi_k(t).$$

The effective value of $K$ is governed by the time–bandwidth product $\Omega\Theta/\pi$, which corresponds to the number of samples determined by the Nyquist interval for a regularly digitized signal with the digitization period $T = \pi/\Omega$. This ensures *normal resolution* capability while maintaining numerical stability. Beyond this range, $\lambda_k$ decays rapidly, and higher-order terms amply numerical noise.

The solution of (13.1) for an arbitrary signal model may further improve the resolution, robustness under noise, and accuracy, but at the time of the development of the doctoral thesis [9] it was not considered suitable for practical implementation.

## B. Extended Fourier Transform of Discrete-Time Signals

In this subsection, the Extended Discrete-Time Fourier Transform (EDTFT) is developed for nonuniformly and uniformly sampled complex-valued signals. The discrete formulations follow the approaches proposed in [6], [7], where analogous derivations for real-valued sequences are given.

### 1) Solution for Unit-Amplitude Model

Define the $K\times K$ time–band limiting matrix $\mathbf{R}$ with entries:

$$r_{l,k} = \frac{\sin(\Omega(t_k - t_l))}{\pi(t_k - t_l)} \quad \text{or} \quad \frac{\sin(\Omega(k-l)T)}{\pi(k-l)T}, \quad k,l = 0,1,\dots,K\text{-}1,$$

the exponential vector $\mathbf{E}_\omega$ of size $K\times 1$ with entries $e^{-i\omega t_l}$ or $e^{-i\omega lT}$, and the basis vector $\mathbf{A}_\omega$ of the same size containing samples $\alpha(\omega, t_k)$ or $\alpha(\omega, kT)$. Then (17.2) – (17.3) can be compactly written as:

$$\mathbf{R}\mathbf{A}_\omega = \mathbf{E}_\omega, \quad \mathbf{A}_\omega = \mathbf{R}^{-1}\mathbf{E}_\omega. \tag{22}$$

Substituting $\mathbf{A}_\omega$ into the extended transform definitions (7), the inverse transform (8), and the amplitude estimator (14) for the *unit-amplitude model* $S(\omega_0) = 1$ ($|\omega_0| \le \Omega$) yields the EDTFT:

$$F_\alpha(\omega) = \mathbf{x}\mathbf{R}^{-1}\mathbf{E}_\omega, \quad |\omega| \le \Omega, \tag{23.1}$$

$$x_\alpha(t) = \mathbf{x}\mathbf{R}^{-1}\mathbf{E}_t, \quad -\infty < t < \infty, \tag{23.2}$$

$$S_\alpha(\omega) = \frac{\mathbf{x}\mathbf{R}^{-1}\mathbf{E}_\omega}{\mathbf{E}_\omega^H\mathbf{R}^{-1}\mathbf{E}_\omega}, \tag{23.3}$$

where $\mathbf{x} = [x(t_0), \dots, x(t_{K-1})]$ or $[x(0), x(T), \dots, x((K-1)T)]$ denotes the row vector of samples and the $K\times 1$ vector $\mathbf{E}_t$ composed as $e_l = \frac{\sin(\Omega(t-t_l))}{\pi(t-t_l)}$ or $\frac{\sin(\Omega(t-lT))}{\pi(t-lT)}$ for nonuniform or uniform sampling cases, respectively.

If the uniform sampling period equals the Nyquist period $T = \pi/\Omega$, then the matrix $\mathbf{R}$ reduces to the identity matrix $\mathbf{I}_K$. In this case, the EDTFT coincides with the classical DTFT:

$$F_\alpha(\omega) = F_\Theta(\omega) = \mathbf{x}\mathbf{E}_\omega, \tag{24.1}$$

$$S_\alpha(\omega) = \frac{1}{K}\mathbf{x}\mathbf{E}_\omega. \tag{24.2}$$

For nonuniform sampling $\mathbf{R} \ne \mathbf{I}_K$, even if the mean sampling interval satisfies the Nyquist constraint. Nevertheless, formulas (23) produce results that are close to the uniform case and typically improve upon the classical nonuniform DTFT (4.2) in terms of spectral leakage and accuracy.

The nominal frequency resolution of the EDTFT in both sampling regimes is $1/(KT)$ i.e., *normal resolution*. For oversampled signals ($T$ or $T_s < \pi/\Omega$,), the EDTFT can provide higher frequency resolution and improved spectral estimates. Practical limits to this improvement arise from finite numerical precision and non-ideal signal discretization process.

In the idealized limit $K \to \infty$, (sample density tending to continuity over $\Theta$), the discrete operator $\mathbf{R}$ converges to the continuous *Sinc* kernel operator and the EDTFT (23.1) converges to the continuous extended transform (20.1).

## 2) Solution for Arbitrary Signal Model

This subsection extends the least-squares formulation of the EDTFT to the case of an arbitrary signal model $S(\omega_0)$. The basis $\alpha(\omega, t_k)$ or $\alpha(\omega, kT)$ can be obtained by solving the system of linear equations (18). In matrix form:

$$\mathbf{A}_\omega = |S(\omega)|^2\mathbf{R}^{-1}\mathbf{E}_\omega. \tag{25}$$

Substituting (25) into (7), (8), and (14) yields the generalized EDTFT formulas:

$$F_\alpha(\omega) = \mathbf{x}\mathbf{A}_\omega = |S(\omega)|^2\mathbf{x}\mathbf{R}^{-1}\mathbf{E}_\omega, \quad |\omega| \le \Omega, \tag{26.1}$$

$$x_\alpha(t) = \mathbf{x}\mathbf{R}^{-1}\mathbf{E}_t, \quad -\infty < t < \infty, \tag{26.2}$$

$$S_\alpha(\omega) = \frac{\mathbf{x}\mathbf{A}_\omega}{\mathbf{E}_\omega^H\mathbf{A}_\omega} = \frac{\mathbf{x}|S(\omega)|^2\mathbf{R}^{-1}\mathbf{E}_\omega}{\mathbf{E}_\omega^H|S(\omega)|^2\mathbf{R}^{-1}\mathbf{E}_\omega} = \frac{\mathbf{x}\mathbf{R}^{-1}\mathbf{E}_\omega}{\mathbf{E}_\omega^H\mathbf{R}^{-1}\mathbf{E}_\omega}. \tag{26.3}$$

Here, the elements of the matrix **R** and vector $\mathbf{E}_t$ are expressed by integrals

$$r_{l,k} = \frac{1}{2\pi}\int_{-\Omega}^{\Omega}|S(\omega_0)|^2 e^{i\omega_0(t_k-t_l)}d\omega_0 \quad \text{or} \quad r_{l,k} = \frac{1}{2\pi}\int_{-\Omega}^{\Omega}|S(\omega_0)|^2 e^{i\omega_0(k-l)T}d\omega_0, \tag{27.1}$$

$$e_l = \frac{1}{2\pi}\int_{-\Omega}^{\Omega}|S(\omega)|^2 e^{i\omega(t-t_l)}d\omega \quad \text{or} \quad e_l = \frac{1}{2\pi}\int_{-\Omega}^{\Omega}|S(\omega)|^2 e^{i\omega(t-lT)}d\omega, \tag{27.2}$$

for nonuniformly or uniformly sampled signals, respectively.

If $|S(\omega_0)|^2 \approx |S_\alpha(\omega_0)|^2$, then (27.1) approximates the autocorrelation function of **x**. The inverse transform (26.2), evaluated at $t = t_k$ or $t = kT$, reproduces the original sequence **x** since $\mathbf{E}_t = \mathbf{R}$.

The frequency resolution of the EDTFT is inversely proportional to $|S(\omega)|^2\mathbf{E}_\omega^H\mathbf{R}^{-1}\mathbf{E}_\omega = F_\alpha(\omega)/\,S_\alpha(\omega)$ and varies across the frequency range $|\omega| \le \Omega$.

## 3) Iterative EDTFT Algorithm

In most cases, the exact signal model spectrum $S(\omega_0)$ is not known. However, the amplitude spectrum obtained with the *unit-amplitude model* can serve as an initial approximation. This motivates an iterative procedure proposed in **[6]**, where the model spectrum gradually converges to the signal spectrum $S_\alpha(\omega)$.

The procedure is as follows:

*Iteration 1:* Compute $S_\alpha^{(1)}(\omega)$ using (23.3) by applying *unit-amplitude model*, $S(\omega_0) = 1$.

*Iteration 2:* Compute $S_\alpha^{(2)}(\omega)$ using (26.3) with $S(\omega_0) = S_\alpha^{(1)}(\omega_0)$.

*Iteration 3:* Compute $S_\alpha^{(3)}(\omega)$ using (26.3) with $S(\omega_0) = S_\alpha^{(2)}(\omega_0)$.

…

*Iteration it:* Compute $S_\alpha^{(it)}(\omega)$ using (26.3) with $S(\omega_0) = S_\alpha^{(it-1)}(\omega_0)$.

The iterations continues either:

- the maximum number of iterations is reached, or
- the power spectrum converges, i.e., $\left|S_\alpha^{(it)}(\omega)\right|^2 \approx \left|S_\alpha^{(it-1)}(\omega)\right|^2$.

The final EDTFT output $F_\alpha(\omega)$ is computed from (26.1) using the last successful iteration result.

By default, the *unit-amplitude model* is used in the first iteration. However, if prior information about the signal characteristics is available, it can be incorporated to define a more realistic model, thus reducing the number of required iterations and improving convergence rate.

# IV. Extended Discrete Fourier Transform

The EDTFT defined in the previous section is a continuous function of frequency over the interval $|\omega| \le \Omega$. In contrast, the Extended Discrete Fourier Transform (EDFT) evaluates the EDTFT on a discrete frequency grid

$$\omega_n \in [-\Omega, \Omega), \quad n = 0,1, \dots, N-1,$$

where $N \ge K$. The number of frequency samples $N$ must be sufficiently large to allow replacement of the continuous integrals in (27.1), used for calculation of the correlation matrix **R** in (25) and (26), with finite sums:

$$r_{l,k} = \frac{1}{2\pi}\int_{-\Omega}^{\Omega} |S(\omega_0)|^2 e^{i\omega_0(t_k - t_l)} d\omega_0 \approx \frac{\Omega}{\pi N}\sum_{n=0}^{N-1} |S(\omega_n)|^2 e^{i\omega_n(t_k - t_l)}, \tag{28.1}$$

$$r_{l,k} = \frac{1}{2\pi}\int_{-\Omega}^{\Omega} |S(\omega_0)|^2 e^{i\omega_0(k-l)T} d\omega_0 \approx \frac{\Omega}{\pi N}\sum_{n=0}^{N-1} |S(\omega_n)|^2 e^{i\omega_n(k-l)T}, \tag{28.2}$$

for $l, k = 0, 1, \dots, K-1$.

The corresponding matrices **R** have the structures:

$$\mathbf{R} = \begin{bmatrix} r_{0,0}(0) & r_{0,1}(t_1 - t_0) & \cdots & r_{0,K-1}(t_{K-1} - t_0) \\ r_{1,0}(t_0 - t_1) & r_{1,1}(0) & & r_{1,K-1}(t_{K-1} - t_1) \\ \vdots & & \ddots & \vdots \\ r_{K-1,0}(t_0 - t_{K-1}) & r_{K-1,1}(t_1 - t_{K-1}) & \cdots & r_{K-1,K-1}(0) \end{bmatrix}, \tag{29.1}$$

$$\mathbf{R} = \begin{bmatrix} r_{0,0}(0) & r_{0,1}(T) & \cdots & r_{0,K-1}((K-1)T) \\ r_{1,0}(-T) & r_{1,1}(0) & & r_{1,K-1}((K-2)T) \\ \vdots & & \ddots & \vdots \\ r_{K-1,0}(-(K-1)T) & r_{K-1,1}(-(K-2)T) & \cdots & r_{K-1,K-1}(0) \end{bmatrix}. \tag{29.2}$$

The matrix **R** always exhibits Hermitian symmetry, $r_{l,k} = r_{k,l}^*$, and in the uniformly sampled case, also a Toeplitz structure. The elements $r_{l,k}$ represent the autocorrelation function and may be computed by applying the Inverse DFT (IDFT) to the model power spectrum $|S(\omega_n)|^2$.

The normalized frequency $\Omega/\pi = 2f_u$ is taken as a unity in EDFT calculations, where $f_u$ is the upper signal frequency. The discrete frequency grid $\{\omega_n\} = \{2\pi f_n\}$ is typically chosen uniformly over $[-f_u, f_u)$, balancing spectral accuracy and computational effort.

For uniform sampling $x(kT)$, the upper frequency or sampling period must be adjusted to satisfy $2f_uT = 1$, ensuring EDFT operation strictly within one Nyquist zone and preventing aliasing. Nonuniform sampling $x(t_k)$ does not require this strict constraint—mean sampling density may span more than one zone ($2f_uT_s \ge 1$). However, the total spectral content of the signal must still not exceed one Nyquist zone.

## A. Non-Iterative EDFT

The non-iterative EDFT of nonuniform or uniform sequences is defined by:

$$\mathbf{R} = \frac{1}{N}\mathbf{EWE}^H, \tag{30.1}$$

$$\mathbf{A} = \mathbf{R}^{-1}\mathbf{EW}, \tag{30.2}$$

$$\mathbf{F} = \mathbf{xA} = \mathbf{xR}^{-1}\mathbf{EW}, \tag{30.3}$$

$$\mathbf{S} = \frac{\mathbf{xA.}}{diag(\mathbf{E}^H\mathbf{A})} = \frac{\mathbf{xR}^{-1}\mathbf{E.}}{diag(\mathbf{E}^H\mathbf{R}^{-1}\mathbf{E})}. \tag{30.4}$$

Here:

- **W** is an *N*×*N* diagonal weight matrix containing a rough estimate of the signal power spectrum $|S(f_n)|^2$.
- **R** is the discrete correlation matrix of size *K*×*K*.
- **A** is the Extended Fourier basis of size *K*×*N*.
- **E** is the Fourier matrix of size *K*×*N*, with elements $e^{-i2\pi f_n t_k}$ (nonuniform sampling) or $e^{-i2\pi f_n kT}$ (uniform sampling).

To account for the omnipresence of non-zero noise power in the range $[-f_u, f_u)$, the diagonal entries of **W** are usually set as strictly positive $|S(f_n)|^2 > 0$ for all $n$. This also improves the numerical stability of algorithm.

## B. Reduction to the Classical DFT

With a *unit-amplitude model* **W** = **I**, and a uniform sequence with a matching uniform frequency grid, the EDFT reduces exactly to the classical DFT:

$$\mathbf{R} = \frac{1}{N}\mathbf{EIE}^H = \mathbf{I}_K, \tag{31.1}$$

$$\mathbf{A} = \mathbf{I}_K^{-1}\mathbf{EI} = \mathbf{E}, \tag{31.2}$$

$$\mathbf{F} = \mathbf{xE}, \tag{31.3}$$

$$\mathbf{S} = \frac{\mathbf{xE.}}{diag(\mathbf{E}^H\mathbf{E})} = \frac{\mathbf{xE.}}{\mathbf{1}_N K} = \frac{1}{K}\mathbf{xE.} \tag{31.4}$$

Here, $\mathbf{I}_K$ is the identity matrix of size *K*×*K* and $\mathbf{1}_N$ is a vector of one's of size *1*×*N*.

For nonuniform time samples or nonuniform frequency grids, the Fourier basis is no longer orthogonal, and:

$$\mathbf{R} = \frac{1}{N}\mathbf{EE}^H \neq \mathbf{I}_K, \tag{32.1}$$

$$\mathbf{A} = \left(\frac{1}{N}\mathbf{EE}^H\right)^{-1}\mathbf{E}, \tag{32.2}$$

$$\mathbf{F} = \mathbf{x}\left(\frac{1}{N}\mathbf{EE}^H\right)^{-1}\mathbf{E}, \tag{32.3}$$

$$\mathbf{S} = \frac{\mathbf{xR}^{-1}\mathbf{E.}}{diag(\mathbf{E}^H\mathbf{R}^{-1}\mathbf{E})}. \tag{32.4}$$

Expression (32.3) resembles the Ordinary Least Squares (OLS) solution normalized by 1/*N*.

### Frequency Resolution

The classical DFT yields constant frequency resolution, since $diag(\mathbf{E}^H\mathbf{E}) = K$ for all frequencies in (31.4) and $diag(\mathbf{E}^H\mathbf{R}^{-1}\mathbf{E}) \approx K$ in (32.4) for the OLS solution. The resolution of the EDFT varies with frequency according to $diag(\mathbf{E}^H\mathbf{A}) \in (0, N]$ in formula (30.4), but, like the classical DFT, it must satisfy *global resolution constraint*:

$$\sum diag(\mathbf{E}^H\mathbf{A}) = \sum \mathbf{F./S} = NK. \tag{33}$$

Thus, *high resolution* at some frequencies requires reduced resolution elsewhere.

## C. Iterative EDFT Algorithm

The iterative EDFT is defined by the following update equations:

$$\mathbf{R}^{(it)} = \frac{1}{N}\mathbf{E}\mathbf{W}^{(it)}\mathbf{E}^{H}, \tag{34.1}$$

$$\mathbf{F}^{(it)} = \mathbf{x}\left(\mathbf{R}^{(it)}\right)^{-1}\mathbf{E}\mathbf{W}^{(it)}, \tag{34.2}$$

$$\mathbf{S}^{(it)} = \frac{\mathbf{x}\left(\mathbf{R}^{(it)}\right)^{-1}\mathbf{E}.}{diag(\mathbf{E}^{H}(\mathbf{R}^{(it)})^{-1}\mathbf{E})'}, \tag{34.3}$$

$$\mathbf{W}^{(it+1)} = diag\left(\left|\mathbf{S}^{(it)}\right|^{2}\right), \tag{34.4}$$

with iteration index $it = 1, 2, \ldots, I$. The initial weight matrix is set to $\mathbf{W}^{(1)} = \mathbf{I}$. If a different initial weight is used, it must contain at least $K$ non-zero diagonal elements to maintain numerical stability.

### Convergence Criteria

Iterations may be terminated early if either of the following conditions is satisfied:

- Spectral change becomes sufficiently small:

$$\frac{|sum(\mathbf{W}^{(it+1)}) - sum(\mathbf{W}^{(it)})|}{sum(\mathbf{W}^{(2)})} < \epsilon_c. \tag{35}$$

- Resolution constraint (33) is violated:

$$\left|1 - \frac{sum(\mathbf{F}./\mathbf{S})}{NK}\right| > \epsilon_r. \tag{36}$$

Here, the thresholds $\epsilon_c$ and $\epsilon_r$ typically take values around $10^{-4}$. In practice, the number of iterations required to reach convergence is usually between 10 and 15.

### Special Case: $N = K$

When the number of frequency samples equals the number of time samples ($N = K$), the EDFT becomes non-iterative and reduces to the classical DFT (31) or the OLS formulation (32). In this case, the weighting matrix has no further effect **[8]**.

## D. Inverse EDFT

Applying Inverse Extended Discrete Fourier Transform (IEDFT) to the spectral coefficients $\mathbf{F}$ of any iteration yields:

$$\mathbf{x} = \frac{1}{N}\mathbf{F}\mathbf{E}^{H}, \tag{37}$$

where $\mathbf{E}^{H}$ denotes the Hermitian transpose of Fourier matrix.

For case $N > K$, the algorithm reconstructs an extrapolated sequence:

$$\mathbf{x}_{\alpha} = \frac{1}{N}\mathbf{F}\mathbf{E}_{N}^{H}, \tag{38}$$

where $\mathbf{E}_{N}^{H}$ is an $N\times N$ exponential matrix with entries $e^{i2\pi f_n t_m}$ or $e^{i2\pi f_n mT}$, for $n, m = 0, 1, \ldots, N-1$.

Thus, $\mathbf{x}_{\alpha}$ represents the original signal extended through forward and backward extrapolation, and through interpolation in cases where samples are missing.

Importantly, the frequency resolution of the EDFT is fundamentally governed by the number of frequency samples $N$, rather than the number of time samples $K$. Consequently, the EDFT offers the theoretical potential to improve local frequency resolution by a factor of $N/K$ relative to the classical DFT. This improvement is an upper bound and is not guaranteed to be realized at all frequencies of interest, as there is a *global resolution constraint* defined in (33).

## E. Properties of EDFT

This subsection summarizes several important properties of the EDFT that complement the results discussed earlier. The EDFT is applicable to both uniform and nonuniform input/output sequences, and several of its features depend on this distinction.

### Connection with DFT and OLS

The similarity between EDFT, the classical DFT, and Ordinary Least Squares (OLS) becomes apparent when the corresponding expressions are written using identity matrices:

$$\begin{aligned}
&\text{EDFT: } \mathbf{xA} &&= \mathbf{xR}^{-1}\mathbf{EW} &&= \mathbf{x}\left(\frac{1}{N}\mathbf{EWE}^H\right)^{-1}\mathbf{EW},\\
&\text{DFT: } \mathbf{xE} &&= \mathbf{xI}_K^{-1}\mathbf{EI} &&= \mathbf{x}\left(\frac{1}{N}\mathbf{EIE}^H\right)^{-1}\mathbf{EI},\\
&\text{OLS: } &&\mathbf{x}\left(\frac{1}{N}\mathbf{EE}^H\right)^{-1}\mathbf{E} &&= \mathbf{x}\left(\frac{1}{N}\mathbf{EIE}^H\right)^{-1}\mathbf{EI}.
\end{aligned} \tag{39}$$

The right-hand expressions coincide for DFT and OLS; EDFT differs only in the diagonal weighting matrix $\mathbf{W}$, whose elements may be unequal. By default, the first EDFT iteration uses $\mathbf{W} = \mathbf{I}$. This means that the iterative EDFT starts as a classical DFT for uniform input/output. Since under nonuniform sampling conditions $\frac{1}{N}\mathbf{EE}^H \neq \mathbf{I}_K$, the first iteration coincides with OLS.

A noteworthy special case arises when the input length $K = N$. Then the Fourier matrix $\mathbf{E}$ is square, and the identity $\mathbf{E}(\frac{1}{N}\mathbf{E}^H\mathbf{E})^{-1} = (\frac{1}{N}\mathbf{EE}^H)^{-1}\mathbf{E}$ holds. Multiplying $\mathbf{R} = \frac{1}{N}\mathbf{EWE}^H$ from the right by $\mathbf{E}(\frac{1}{N}\mathbf{E}^H\mathbf{E})^{-1}$, we obtain:

$$\mathbf{RE}(\frac{1}{N}\mathbf{E}^H\mathbf{E})^{-1} = \mathbf{EW}\left(\frac{1}{N}\mathbf{E}^H\mathbf{E}\right)\left(\frac{1}{N}\mathbf{E}^H\mathbf{E}\right)^{-1} = \mathbf{EW},$$

which yields $\mathbf{EW} = \mathbf{R}(\frac{1}{N}\mathbf{EE}^H)^{-1}\mathbf{E}$ . Substituting into (39) gives:

$$\mathbf{xR}^{-1}\mathbf{EW} = \mathbf{xR}^{-1}\mathbf{R}(\frac{1}{N}\mathbf{EE}^H)^{-1}\mathbf{E} = \mathbf{x}(\frac{1}{N}\mathbf{EE}^H)^{-1}\mathbf{E}. \tag{40}$$

Thus, when $K = N$, the EDFT output does not depend on $\mathbf{W}$ and can be computed non-iteratively using OLS, or—when the sampling is uniform—by the classical DFT since $\frac{1}{N}\mathbf{EE}^H = \mathbf{I}_K$.

### Inverse of EDFT, DFT and OLS

The IEDFT is computed using (37) and is implemented by applying the Hermitian transpose of the Fourier basis $\frac{1}{N}\mathbf{E}^H$ to the output (39). It exactly reconstructs the input sequence:

$$\begin{aligned}
&\text{Inverse of EDFT: } &&\mathbf{x}\left(\frac{1}{N}\mathbf{EWE}^H\right)^{-1}\left(\frac{1}{N}\mathbf{EWE}^H\right) = \mathbf{x},\\
&\text{Inverse of DFT/OLS: } &&\mathbf{x}\left(\frac{1}{N}\mathbf{EIE}^H\right)^{-1}\left(\frac{1}{N}\mathbf{EIE}^H\right) = \mathbf{x}.
\end{aligned} \tag{41}$$

Therefore, the same inverse basis $\frac{1}{N}\mathbf{E}^H$ recovers the input sequence from the output of EDFT, DFT, or OLS.

A distinctive EDFT feature is its ability to reconstruct extrapolated sequences (38) via the extended basis $\frac{1}{N}\mathbf{E}_N^H$ of size $N{\times}N$. Splitting the inverse basis into $\left[\frac{1}{N}\mathbf{E}^H\ \frac{1}{N}\mathbf{E}_{(N-K)}^H\right]$, we obtain:

$$\mathbf{x}_\alpha = \mathbf{x}\left(\frac{1}{N}\mathbf{EWE}^H\right)^{-1}\frac{1}{N}\mathbf{EWE}_N^H = \left[\mathbf{x}\ \ \mathbf{x}\left(\frac{1}{N}\mathbf{EWE}^H\right)^{-1}\frac{1}{N}\mathbf{EWE}_{(N-K)}^H\right]. \tag{42}$$

In this way, the IEDFT reconstructs the original sequence (the first $K$ samples) and appends an extrapolated portion. Consequently, the extended reconstruction is band-limited to the selected frequency set.

Applying IDFT to the classical DFT output (39) yields:

$$\mathbf{x}\left(\frac{1}{N}\mathbf{EIE}^{H}\right)^{-1}\frac{1}{N}\mathbf{EIE}_{N}^{H}=\left[\mathbf{x}\ \mathbf{x}\left(\frac{1}{N}\mathbf{EIE}_{(N-K)}^{H}\right)\right]=\left[\mathbf{x}\ \mathbf{0}_{(N-K)}\right],$$

i.e., the original sequence padded with zeros.

The inverse OLS transform corresponds to (42) with $\mathbf{W}=\mathbf{I}$: it reproduces the input and adds a rapidly vanishing extrapolated part. Hence, the choice $\mathbf{W}\neq\mathbf{I}$ is essential for the distinct behavior of EDFT.

A further practical use of EDFT is interpolation and resampling: applying the IEDFT with a uniform time grid in $\mathbf{E}_{N}^{H}$ produces uniformly sampled output from nonuniform input.

## Orthogonality

The EDFT basis matrix $\mathbf{A}$ and the Hermitian transpose of the Fourier basis $\mathbf{E}^{H}$ constitute a biorthogonal or dual pair satisfying

$$\mathbf{AE}^{H}=N\mathbf{I}_{K},\text{ but generally }\mathbf{E}^{H}\mathbf{A}\neq N\mathbf{I}_{N}. \tag{43}$$

This property follows directly from (41) and guarantees exact reconstruction of the input sequence via the IEDFT. It is important to emphasize condition (43) holds for both uniform and nonuniform input/output configurations. In contrast, for the classical DFT, orthogonality is preserved only under uniform conditions.

## Parseval–Plancherel Theorem

Let $\mathbf{F}_{x}$ and $\mathbf{F}_{y}$ are $N$-point EDFTs of sequences $\mathbf{x}$ and $\mathbf{y}$ (both of length $K$), and let $\mathbf{x}_{\alpha}$ and $\mathbf{y}_{\alpha}$ be their IEDFT reconstructions. Then:

$$\mathbf{xy}^{H}\leq\mathbf{x}_{\alpha}\mathbf{y}_{\alpha}^{H}=\frac{1}{N}\mathbf{F}_{x}\mathbf{F}_{y}^{H},\quad \mathbf{xx}^{H}\leq\mathbf{x}_{\alpha}\mathbf{x}_{\alpha}^{H}=\frac{1}{N}\mathbf{F}_{x}\mathbf{F}_{x}^{H},\quad K\leq N. \tag{44}$$

Since, $\mathbf{y}_{\alpha}^{H}=\frac{1}{N}\mathbf{EF}_{y}^{H}$ and $\mathbf{F}_{x}=\mathbf{x}_{\alpha}\mathbf{E}$, we obtain $\mathbf{x}_{\alpha}\mathbf{y}_{\alpha}^{H}=\frac{1}{N}\mathbf{x}_{\alpha}\mathbf{EF}_{y}^{H}=\frac{1}{N}\mathbf{F}_{x}\mathbf{F}_{y}^{H}$.

Plancherel's theorem is a special case of $\mathbf{y}=\mathbf{x}$. The result holds exactly for uniform input/output and approximately for nonuniform EDFT. If the nonuniform output is first resampled onto a uniform grid using the IEDFT, then the identity becomes exact.

## Linearity

For two sequences $\mathbf{x}_1$ and $\mathbf{x}_2$, and scalars $a,b\in\mathrm{C}$, let $\mathbf{x}=a\mathbf{x}_1+b\mathbf{x}_2$. Using (39), we have

$$\mathbf{xA}=(a\mathbf{x}_1+b\mathbf{x}_2)\mathbf{A}=a\mathbf{x}_1\mathbf{A}+b\mathbf{x}_2\mathbf{A}. \tag{45}$$

This equality is exact for Non-Iterative EDFT or Iterative EDFT when a *single* EDFT basis $\mathbf{A}$ (obtained for the composite sequence $\mathbf{x}$) is used. If iterative algorithm is applied separately to $\mathbf{x}_1$ and $\mathbf{x}_2$, yielding bases $\mathbf{A}_1$ and $\mathbf{A}_2$, then:

$$\mathbf{xA}\approx a\mathbf{x}_1\mathbf{A}_1+b\mathbf{x}_2\mathbf{A}_2. \tag{46}$$

Nevertheless, applying the IEDFT (37) to either side of (46) yields the same result, $\mathbf{x}=a\mathbf{x}_1+b\mathbf{x}_2$. The approximation arises because the extrapolated portion of the sequences does not match, and this can be explained by the adaptive nature of EDFT in the frequency domain.

## Time and Frequency Reversal

The EDFT exhibits the same reversal properties as the DFT. If $\mathbf{x}_{(K-k)}$ denotes the time-reversed version of the sequence $\mathbf{x}$, then the EDFT satisfies

$$\mathbf{F}=\mathbf{xA},\qquad \mathbf{F}_{(N-n)}=\mathbf{x}_{(K-k)}\mathbf{A}, \tag{47}$$

where $\mathbf{F}_{(N-n)}$ is the frequency-reversed version of $\mathbf{F}$. Both EDFTs are evaluated at the same iteration number, ensuring consistency of the adaptive basis.

### Complex Conjugate in Time

The EDFT of a complex conjugate sequence $\mathbf{x}^*$ is equal to the conjugated, frequency-reversed EDFT of the original sequence $\mathbf{x}$:

$$\mathbf{F} = \mathbf{xA}, \qquad \mathbf{F}^*_{(N-n)} = \mathbf{x}^*\mathbf{A}. \tag{48}$$

Thus, EDFT preserves the standard conjugate-symmetry relationships known from the DFT.

### Circular Time Shift

A circular time shift of the input sequence produces a linear phase modulation in the EDFT. If $\mathbf{x}_{(N-t_m)}$ is the circularly shifted sequence and vector $\mathbf{E}_{(t_m)} = e^{-i2\pi f_n t_m}$, then

$$\mathbf{F} = \mathbf{xA}, \qquad \mathbf{F}.\mathbf{E}_{(t_m)} = \mathbf{x}_{(N-t_m)}\mathbf{A}. \tag{49}$$

The shift $t_m$ may be non-integer. For $t_m = NT$ (or $NT_s$ in nonuniform case), the sequence is circularly shifted by a whole cycle and coincides with the original sequence, i.e., $\mathbf{x}_{(N-t_m)} = \mathbf{x}$.

### Circular Frequency Shift

Multiplying a uniformly sampled sequence by a linear phase factor $\mathbf{E}_m = e^{-i2\pi mk/N}$ for some integer $m$ produces a circular frequency shift in the EDFT:

$$\mathbf{F} = \mathbf{xA}, \qquad \mathbf{F}_{(N-m)} = (\mathbf{x}.\mathbf{E}_m)\mathbf{A}. \tag{50}$$

This property holds for uniform input/output sequences and mirrors the standard circular frequency-shift behavior of the classical DFT.

# V. EDFT and Other Nonparametric Approaches

In the preceding sections, the EDFT was derived analytically starting from the Fourier integral by exploiting orthogonality properties and solving the associated least-squares error estimators. In this section, the EDFT framework is compared with several well-known nonparametric spectral estimation methods—specifically, the Capon filter, the Generalized Weighted Least Squares (GWLS) formulation, and the High-Resolution Discrete Fourier Transform (HRDFT) introduced by Sacchi, Ulrych, and Walker (1998). We also briefly examine how EDFT-type iterative algorithms may be derived from these approaches.

## A. Capon Filter Approach

The Capon filter—also known as the Minimum Variance (MV) spectral estimator [3], [11], [12], [25]—may be interpreted as the output of a bank of frequency-selective filters, each centered at a particular analysis frequency. For a uniformly sampled sequence $x(kT)$, the filter output is

$$y_\omega(nT) = \sum_{k=0}^{K-1} x\big((n-k)T\big)\, h_\omega(kT) = \mathbf{x}_{(K-k)}\mathbf{h}_\omega, \;\; n = 0,1,2,\dots,$$

where $\mathbf{x}_{(K-k)} = [x(nT), x\big((n-1)T\big), \dots, x\big((n-K+1)T\big)]$ is vector of input samples, and $\mathbf{h}_\omega = [h_\omega(0), h_\omega(T), \dots, h_\omega\big((K-1)T\big)]^T$ is the frequency-dependent filter coefficient vector.

The Capon filter is designed to minimize the output variance

$$\sigma_y^2 = \boldsymbol{\varepsilon}\{|y_\omega(nT)|^2\} = \mathbf{h}_\omega^H \boldsymbol{\varepsilon}\{\mathbf{x}^{\boldsymbol{H}}_{(K-k)}\mathbf{x}_{(K-k)}\}\mathbf{h}_\omega = \mathbf{h}_\omega^H \mathbf{R}_x \mathbf{h}_\omega, \tag{51}$$

subject to a unit-gain constraint at the analysis frequency:

$$H(\omega) = \sum_{k=0}^{K-1} h_\omega(kT)e^{-i\omega kT} = \mathbf{E}_\omega^T\mathbf{h}_\omega = 1, \quad (52.1)$$

$$H(\omega) = \sum_{k=0}^{K-1} h_\omega^*(kT)e^{i\omega kT} = \mathbf{h}_\omega^H\mathbf{E}_\omega^* = 1, \quad (52.2)$$

where $\mathbf{E}_\omega$ is column vector with entries $e^{-i\omega kT}$, and $\boldsymbol{\varepsilon}\{.\}$ denotes expectation.

The matrix $\mathbf{R}_x = \boldsymbol{\varepsilon}\{\mathbf{x}_{(K-k)}^{\boldsymbol{H}}\mathbf{x}_{(K-k)}\}$ is the sample autocorrelation matrix. A biased estimate of the autocorrelation function is

$$r_{xx}(lT) = \frac{1}{K}\sum_{k=0}^{K-l-1} x\big((k+l)T\big)\,x^*(kT), \qquad l = 0, 1, \dots, K-1,$$

and using the symmetry $r_{xx}(-lT) = r_{xx}^*(lT)$, the matrix $\mathbf{R}_x$ takes Toeplitz-Hermitian form

$$\mathbf{R}_x = \begin{bmatrix} r_{0,0}(0) & r_{0,1}(-T) & \cdots & r_{0,K-1}(-(K-1)T) \\ r_{1,0}(T) & r_{1,1}(0) & & r_{1,K-1}(-(K-2)T) \\ \vdots & & \ddots & \vdots \\ r_{K-1,0}((K-1)T) & r_{K-1,1}((K-2)T) & \cdots & r_{K-1,K-1}(0) \end{bmatrix}.$$

The filter coefficients are obtained by minimizing (51) under the constraints (52) using Lagrange multipliers:

$$J = \mathbf{h}_\omega^H\mathbf{R}_x\mathbf{h}_\omega - \mu(\mathbf{E}_\omega^T\mathbf{h}_\omega - 1) - \lambda(\mathbf{h}_\omega^H\mathbf{E}_\omega^* - 1) \to min. \quad (53)$$

Setting the derivatives with respect to $\mathbf{h}_\omega$ and $\mathbf{h}_\omega^H$ to zero leads to the solution

$$\mathbf{h}_\omega = \frac{\mathbf{R}_x^{-1}\mathbf{E}_\omega^*}{\mathbf{E}_\omega^T\mathbf{R}_x^{-1}\mathbf{E}_\omega^*} \quad (54)$$

and the Capon (MV) power estimate becomes

$$P_{Capon}(\omega) = \mathbf{h}_\omega^H\mathbf{R}_x\mathbf{h}_\omega = \frac{1}{\mathbf{E}_\omega^T\mathbf{R}_x^{-1}\mathbf{E}_\omega^*}. \quad (55)$$

To construct an EDFT-like iteration, the sample autocorrelation matrix $\mathbf{R}_x$ may be replaced by transpose of the EDFT matrix,

$$\mathbf{R}^T = \mathbf{E}^*\mathbf{W}\mathbf{E}^T,$$

with $\mathbf{R}$ defined by (29). The diagonal matrix $\mathbf{W}$ (size $N\times N$) contains power estimates at filter outputs. A single-sample estimate at time $nT = 0$ is

$$|y_\omega(0)|^2 = \left|\mathbf{x}_{(K-k)}\mathbf{h}_\omega\right|^2 = \left|\frac{\mathbf{x}_{(K-k)}(\mathbf{R}^T)^{-1}\mathbf{E}_\omega^*}{\mathbf{E}_\omega^T(\mathbf{R}^T)^{-1}\mathbf{E}_\omega^*}\right|^2,$$

where $\mathbf{x}_{(K-k)}$ is the time-reversed input sequence (uniform or nonuniform).

This motivates the following iterative scheme:

$$\mathbf{R}^{T(it)} = \mathbf{E}^*\mathbf{W}^{(it)}\mathbf{E}^T, \quad (56.1)$$

$$\mathbf{S}_{Capon}^{(it)} = \frac{\mathbf{x}_{(K-k)}\left(\mathbf{R}^{T(it)}\right)^{-1}\mathbf{E}^*.}{diag(\mathbf{E}^T(\mathbf{R}^{T(it)})^{-1}\mathbf{E}^*)}, \quad (56.2)$$

$$\mathbf{W}^{(it+1)} = diag\left(\left|\mathbf{S}_{Capon}^{(it)}\right|^2\right), \quad (56.3)$$

with $\mathbf{W}^{(1)} = \mathbf{I}$ and iteration index $it = 1, 2, \dots, I$.

The resulting power estimate $\left|\mathbf{S}_{Capon}^{(it)}\right|^2$ matches the EDFT power spectrum, although the phase differs because it is evaluated from the time-reversed input.

It is important to emphasize:

- Expression (55) is formally grounded in the Capon derivation.
- The iterative form (56), in contrast, relies on heuristic substitutions (e.g., replacing $\mathbf{R}_x$ by $\mathbf{R}^T$) and evaluates power from a single filter output sample, which is not statistically rigorous.
- Thus, the method in (56) is best interpreted as a filter-bank view of the EDFT, analogous to the classical interpretation of the DFT as a set of parallel filters.
- This Capon-based construction does not reproduce the full capabilities of the EDFT, such as estimating the extended Fourier basis **A**, performing signal reconstruction, or enabling broadband extrapolation.

## B. GWLS Approach

The Generalized Weighted Least Squares (GWLS) framework **[3, 16, 19, 35]** can be applied to spectral estimation using the linear data model

$$\mathbf{x}^T = \mathbf{E}_\omega^* S_{GWLS}(\omega) + \mathbf{e}_Q, \tag{57}$$

with $\mathbf{e}_Q$ represents noise and interference, while the term $\mathbf{E}_\omega^* S_{GWLS}(\omega)$ models the signal component at frequency $\omega$ with unknown complex amplitude $S_{GWLS}(\omega)$. Under this model, GWLS minimizes the quadratic form

$$[\mathbf{x}^T - \mathbf{E}_\omega^* S_{GWLS}(\omega)]^H \, \mathbf{Q}^{-1} [\mathbf{x}^T - \mathbf{E}_\omega^* S_{GWLS}(\omega)],$$

leading to the solution

$$S_{GWLS}(\omega) = \frac{\mathbf{E}_\omega^T \mathbf{Q}^{-1} \mathbf{x}^T}{\mathbf{E}_\omega^T \mathbf{Q}^{-1} \mathbf{E}_\omega^*}, \tag{58}$$

where **Q** (size *K*×*K*) is the covariance matrix of term $\mathbf{e}_Q$.

Two special cases arise naturally:

- Weighted Least Squares (WLS): when all off-diagonal elements of **Q** are zero.
- Ordinary Least Squares (OLS): when $\mathbf{e}_Q$ is assumed to be white noise, giving **Q** = **I**.

A practical difficulty with GWLS is that the noise covariance matrix **Q** is generally unknown and must be estimated from the data simultaneously with $S_{GWLS}(\omega)$. A reasonable starting point is the OLS estimate, i.e., (58) with **Q** = **I**, which serves as the first iteration.

To construct an iterative estimator with EDFT-like behavior, one may substitute the covariance matrix with

$$\mathbf{Q} = \mathbf{R}^T = \mathbf{E}^* \mathbf{W} \mathbf{E}^T,$$

assuming that $\mathbf{W} = diag(|S_{GWLS}(\omega)|^2)$. Under this substitution, expression (58) becomes identical to the EDTFT estimator (26.3):

$$S_{GWLS}(\omega) = \frac{\mathbf{E}_\omega^T (\mathbf{R}^T)^{-1} \mathbf{x}^T}{\mathbf{E}_\omega^T (\mathbf{R}^T)^{-1} \mathbf{E}_\omega^*} = \frac{\mathbf{x} \mathbf{R}^{-1} \mathbf{E}_\omega}{\mathbf{E}_\omega^H \mathbf{R}^{-1} \mathbf{E}_\omega} = S_\alpha(\omega). \tag{59}$$

In that way, the GWLS formulation can be used to iteratively refine the amplitude spectrum in the same manner as the EDFT.

However, several limitations must be highlighted:

- The substitution $\mathbf{Q} = \mathbf{R}^T$ is not directly supported by the GWLS model. In (57), $\mathbf{Q}$ describes only the covariance of the noise component $\mathbf{e}_Q$, whereas $\mathbf{R}^T$ is computed from the *entire* signal $\mathbf{x}^T$, which includes both noise and the signal component $\mathbf{E}_\omega^* S_{GWLS}(\omega)$.
- Consequently, $S_{GWLS}(\omega)$ obtained from (59) contains contributions from all components of the data, including noise, and cannot be interpreted as a pure spectral estimate of the signal term in (57).
- The reconstruction mechanism is inconsistent with the GWLS model. The time-domain signal is restored via the IEDFT,
$$\mathbf{x}^T = \frac{1}{N}\mathbf{E}^*\mathbf{F} \neq \mathbf{E}^*\mathbf{S},$$
whereas model (57) assumes a direct representation $\mathbf{x}^T = \mathbf{E}^*\mathbf{S}$. Using an estimate $S_{GWLS}(\omega) = S_\alpha(\omega)$ in (57) forces an artificial decomposition:
$$\mathbf{e}_Q = \frac{1}{N}\mathbf{E}^*\mathbf{F} - \mathbf{E}^*\mathbf{S},$$
which does not correspond to the original model assumptions. Thus, there is an inherent discrepancy between the GWLS model prediction and the result obtained from (59).

Despite these limitations, the GWLS interpretation remains useful conceptually—particularly in scenarios where the noise spectrum is spread across the entire frequency band while the signal spectrum is localized. In such cases, the EDFT output can be viewed heuristically through the lens of the GWLS data model.

## C. HRDFT Approach

The third nonparametric approach considered here is the High-Resolution Discrete Fourier Transform (HRDFT), originally proposed by Sacchi, Ulrych, and Walker **[10]**. HRDFT provides an iterative Bayesian framework for spectral estimation, producing a high-resolution Fourier spectrum in a manner conceptually similar to the EDFT. The HRDFT algorithm may be expressed through the following iterative procedure:

$$\mathbf{R}^{(it)} = \frac{1}{N}\mathbf{E}\mathbf{W}^{(it)}\mathbf{E}^H, \tag{60.1}$$

$$\mathbf{F}_{HRDFT}^{(it)} = \mathbf{x}\left(\mathbf{R}^{(it)}\right)^{-1}\mathbf{E}\mathbf{W}^{(it)}, \tag{60.2}$$

$$\mathbf{W}^{(it+1)} = diag\left(\left|\frac{1}{N}\mathbf{F}_{HRDFT}^{(it)}\right|^2\right), \tag{60.3}$$

where $it = 1, 2, \ldots, I$ denotes the iteration index and the initial weight matrix is $\mathbf{W}^{(1)} = \mathbf{I}$.

Analogous to the EDFT framework, applying the IDFT (37) to the output of any arbitrary HRDFT iteration (60.2) yields an exact, distortion-free reconstruction of the input sequence $\mathbf{x}$. However, a key distinction lies in the fact that HRDFT lacks an explicit closed-form expression for the amplitude spectrum analogous to (34.3). Instead, the algorithm recursively updates the weights $\mathbf{W}^{(it+1)}$ utilizing the Fourier-domain estimate from the previous iteration.

According to the EDFT data model (11), the corresponding continuous-frequency Fourier transform (12) incorporates a Dirac delta distribution. Consequently, enforcing such an optimization objective introduces severe numerical ill-conditioning and computational instability, causing the operational trajectory and asymptotic convergence properties of the algorithm to deviate substantially from those of the EDFT.

The HRDFT iteration converges toward a representation in which the *entire* signal $\mathbf{x}$ is modeled using approximately $K$ dominant frequency components, whereas the residual spectral power across the remaining $N-K$ bins becomes negligible. Crucially, each resolved line spectrum achieves the maximum physical resolution bound dictated by the total frequency grid dimension

*N*. In this sense, HRDFT behaves like a sparse spectral estimator, concentrating energy into a small set of dominant components.

Importantly, HRDFT—like DFT and EDFT—obeys the *global resolution constraint*

$$\sum(\text{resolution}^{-1}) = NK,$$

and therefore, redistributes resolution across the spectrum rather than increasing it globally. This redistribution mechanism is central to the adaptive resolution behavior observed in HRDFT and EDFT.

## D. Summary of Comparisons

The three nonparametric approaches reviewed in this section—Capon filtering, GWLS estimation, and HRDFT—share conceptual similarities with the EDFT but differ substantially in motivation, mathematical formulation, and practical capability.

The **Capon filter** (Minimum Variance method) constructs a bank of adaptive filters whose outputs estimate spectral power with unity gain at the analysis frequency. When its covariance matrix is replaced by the EDFT correlation matrix, an EDFT-like iterative scheme emerges; however, this modification is *ad hoc* and departs from the theoretical foundations of the Capon method, which is strictly defined for power estimation. As a result, the Capon-based interpretation yields only the power spectrum, lacks a Fourier transform estimate, and cannot reconstruct the time-domain signal. Its iterative form is therefore best viewed as a filter-bank interpretation of EDFT rather than a full analogue.

The **GWLS method** provides a generalized weighted least-squares formulation for estimating complex amplitudes at individual frequencies, using a noise covariance matrix. When this covariance matrix is replaced by the EDFT correlation matrix, the resulting estimator reproduces the EDFT amplitude spectrum. However, this substitution conflicts with the GWLS data model, in which the covariance matrix should describe only the noise term and not the full signal. Consequently, the adapted GWLS formulation captures only part of the EDFT behavior: it reproduces the spectrum but does not conform to the model's assumptions, cannot separate signal and noise components consistently, and cannot reproduce the EDFT-based reconstruction.

The **HRDFT algorithm**, like EDFT, iteratively adapts a weighting matrix to obtain high-resolution spectral estimates. Its formulation is more closely aligned with EDFT than the two previous approaches, yet it still lacks an explicit amplitude estimator and therefore converges to a representation dominated by *K* high-energy spectral components, with minimal power elsewhere. HRDFT satisfies the *global resolution constraint* but differs from EDFT in both convergence behavior and the resulting spectral distribution.

In summary, although all three approaches share the idea of iteratively refining spectral estimates, only EDFT provides a complete and self-consistent framework: it yields the Fourier transform, amplitude, and power spectra; preserves the input sequence through exact inverse transformation; supports both uniform and nonuniform sampling; and enables controlled extrapolation and enhanced frequency resolution. Capon filtering and GWLS partially resemble EDFT under specific substitutions, whereas HRDFT can achieve higher resolution at the expense of stability and is best suited for estimating line spectra. Hence, EDFT may be regarded as a more general framework that unifies and extends the strengths of these nonparametric approaches.

# VI. Computer simulations

All numerical experiments were performed using MATLAB software available online to registered users. The classical DFT for uniformly sampled sequences on a uniform frequency grid was computed using MATLAB's built-in Fast Fourier Transform (FFT) routine **[2]**. For nonuniformly sampled sequences or nonuniform frequency grids, the Nonuniform Fast Fourier Transform (NUFFT) was employed **[4]**, also available as a native MATLAB function. EDFT results were generated using the implementation provided in **APPENDIX D**, while the HRDFT spectra were obtained by running the EDFT algorithm with weighting matrix **W** proportional to the magnitude of the Fourier-domain estimate. Time-domain reconstruction was performed using the Inverse FFT (IFFT) when the frequency grid was uniform, and using the IEDFT listed in **APPENDIX E** when the frequency grid was nonuniform.

Throughout Figures 1–10, the true spectrum is plotted in red, the FFT/NUFFT in blue, and the EDFT estimate in black.

## A. EDFT vs. FFT and NUFFT for Composite Complex-Valued Sequence

To validate the proposed EDFT algorithm, simulations were performed using the test signal from **[9]**. The reference (true) spectrum consists of three components within the normalized frequency range [−0.5, 0.5) Hz:

1. A band-limited random component in the interval [−0.5, −0.25] Hz,
2. A rectangular pulse occupying [0, 0.25] Hz, and
3. Two deterministic unit-amplitude complex exponentials at 0.35 Hz and 0.3985 Hz.

These three components represent, respectively, the random, transient, and deterministic contributions of a composite signal with upper frequency $f_u = 0.5$ Hz.

Uniform and nonuniform sequences of length $K = 64$ were generated using a simulated 10-bit analog-digital converter (ADC). The uniform sampling period was $T = 1$ s. For the nonuniform case, sampling instants were generated as $t_k = kT + \tau_k$, where the perturbations $\tau_k$ were independent random variables uniformly distributed over [0, 0.8] s. In both cases, the ADC introduced a noise floor of approximately –60 dB.

Figure 1 evaluates the performance of EDFT for a uniformly sampled sequence of length $K = 64$ using $N = 1000$ analysis frequencies, corresponding to a DFT bin spacing of $2f_u/N = 0.001$ Hz.

**Fig. 1(a)**—The non-iterative EDFT power spectrum, computed as $10\log_{10}(|\mathbf{S}|^2)$, uses the true spectrum (red line with strictly positive values in range [−0.5, 0.5) Hz) as the initial weight matrix **W**. Because the weighting is ideal, the estimate nearly coincides with the 15-iteration EDFT result.

**Fig. 1(b)**—EDFT power spectrum after 15 iterations with **W** = **I**, confirming convergence of the iterative algorithm. All signal components—random, transient, and deterministic—are accurately reconstructed.

**Fig. 1(c)**—EDFT power spectral density (PSD) estimate, computed as $10\log_{10}(|\mathbf{F}|^2/N)$. Sharp, well-localized peaks appear at the deterministic frequencies, in contrast with the broader peaks produced by the FFT, whose half-width corresponds approximately to the *normal resolution*.

**Fig. 1(d)**—Relative frequency resolution, computed as $\frac{1}{2f_uTK}\mathbf{F}./\mathbf{S}$. EDFT enhances resolution (values > 1) near strong components and reduces it (values < 1) in noise-only regions, while preserving the global resolution budget (equal area under EDFT and FFT curves). With $N/K \approx 1000/64 \approx 16$, EDFT achieves up to 16× improvement in resolving power. The peak resolution occurs near the tone 0.35 Hz. The tone 0.3985 Hz does not reach the maximum resolution because

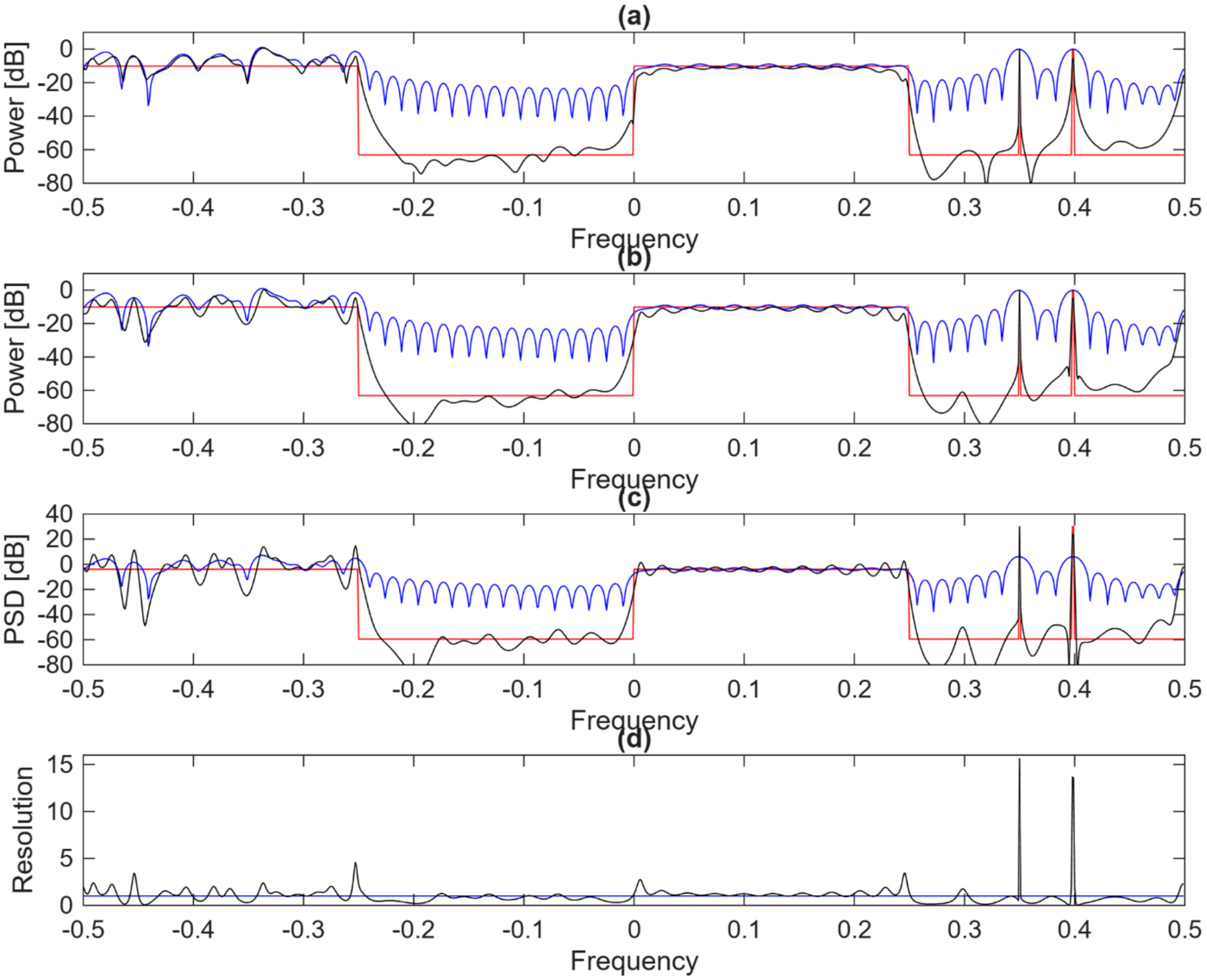

**Figure 1 – EDFT and FFT performance for a uniformly sampled sequence.**

**(a)** Power spectrum of FFT and Non-iterative EDFT computed using the true spectrum as weighting matrix **W**;
**(b)** Power spectrum FFT and EDFT after 15 iterations initialized with **W** = **I**;
**(c)** EDFT power spectral density (PSD) estimate compared with FFT;
**(d)** Relative frequency resolution of EDFT versus FFT, demonstrating localized resolution enhancement at deterministic components.

it does not lie exactly on the EDFT grid (0.001 Hz spacings), spreading its energy across adjacent bins—a well-known DFT effect that disappears with a finer frequency grid.

Figure 2 illustrates the performance of EDFT applied to a nonuniformly sampled 64-point sequence, using the same analysis parameters as in Figure 1. Although the sampling instants are perturbed by uniformly distributed timing deviations, EDFT maintains high spectral accuracy and remains robust to sampling irregularities.

**Fig. 2(a)** and **Fig.2(b)**—As in the uniformly sampled case, the EDFT spectrum computed non-iteratively (using the true spectral weights) is nearly indistinguishable from the 15-iteration EDFT result. This demonstrates stable convergence of the algorithm even under nonuniform sampling. In contrast, NUFFT accuracy degrades noticeably when timing perturbations are introduced.

**Fig. 2(c)**—The PSD of the nonuniform EDFT output closely matches its uniform-sampling counterpart, confirming the robustness of EDFT to timing jitter. The NUFFT estimate, however, exhibits visible degradation across all spectral components.

**Fig. 2(d)**—Relative frequency resolution is computed as $\frac{1}{2f_u T_s K}\mathbf{F}./\mathbf{S}$ using the mean sampling period $T_s$ = 1 s. Since $2f_uT_s = 1$, the nonuniform sequence is processed within a single Nyquist

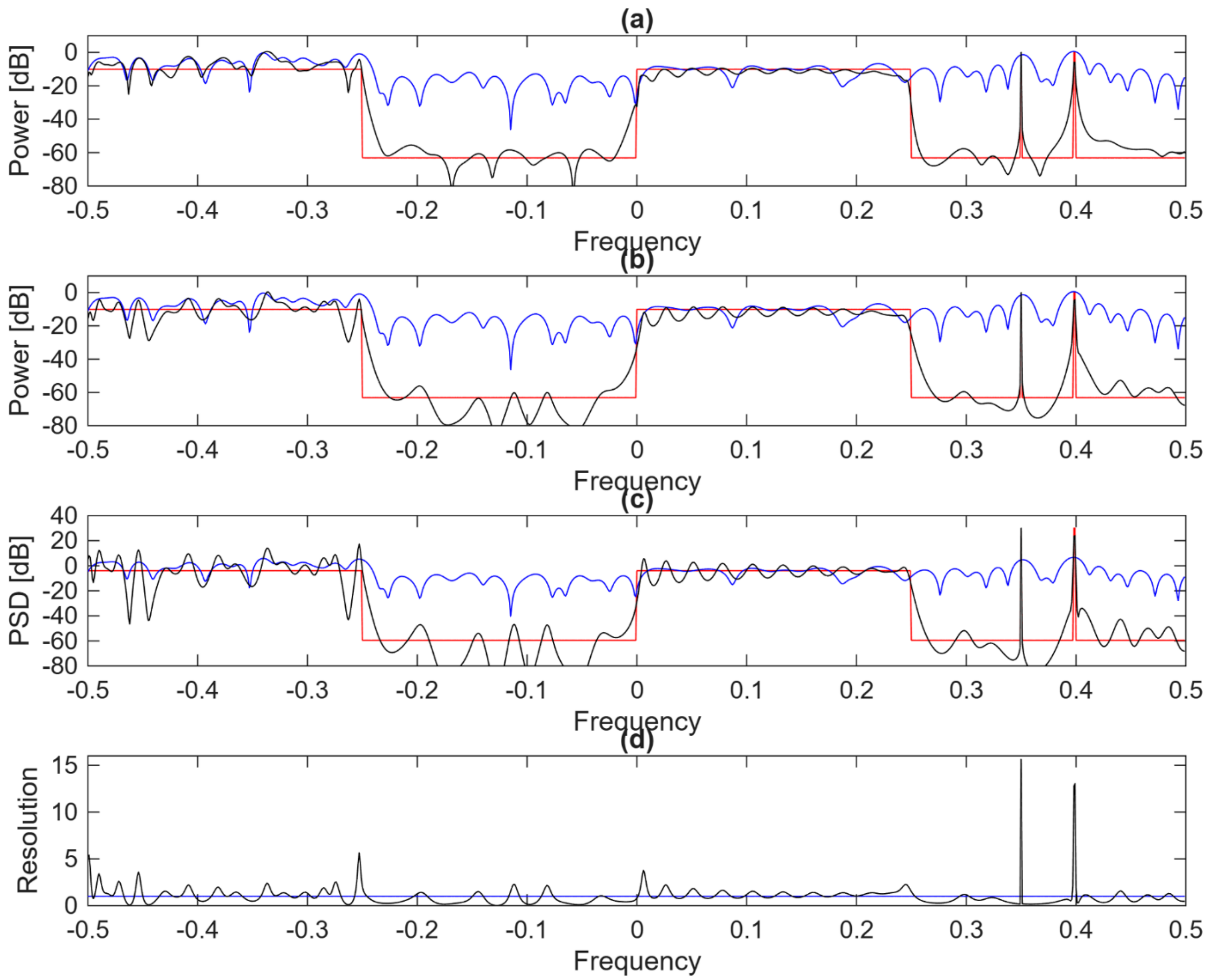

**Figure 2 – EDFT and NUFFT performance for nonuniformly sampled sequence.**

**(a)** Power spectrum of NUFFT and non-iterative EDFT using true spectral weights;
**(b)** Power spectrum of NUFFT and EDFT after 15 iterations with **W** = **I**;
**(c)** PSD comparison of NUFFT and EDFT for nonuniform input;
**(d)** Relative frequency resolution of EDFT versus NUFFT, demonstrating adaptive sharpening and resolution redistribution.

zone. The EDFT resolution behavior mirrors that of Fig. 1(d), confirming that EDFT consistent high-quality spectral estimation for both uniform and nonuniform sampling patterns.

Figure 3 investigates the behavior of EDFT and compare with NUFFT when the frequency grid itself is nonuniform. The uniform FFT grid is perturbed by adding random jitter within ±0.0004 Hz, producing a randomized set of frequency evaluation points.

**Fig. 3(a)**—PSD of the uniform time-domain sequence on the perturbed frequency grid. The EDFT accurately recovers the expected spectral structure, closely resembling the uniform-grid result of Fig. 1(c). Differences appear only in the fine placement of peaks, attributable to the jittered grid and its local frequency irregularities. NUFFT performs similarly to FFT in Fig. 1(c).

**Fig. 3(b)**—The EDFT relative resolution adapts to the unequal spacing of frequency samples. The peak corresponding to the 0.3985 Hz tone lies closer to an irregular grid point than in Fig. 1(c), while the peak at 0.35 Hz no longer aligns with a grid point. This causes the resolution redistribution to adjust accordingly, demonstrating EDFT's sensitivity to the underlying sampling geometry.

**Fig. 3(c)** and **Fig. 3(d)**—Analysis of the nonuniform time-domain sequence on a nonuniform frequency grid yields results consistent with those in (a)–(b). EDFT remains stable and accurate despite irregularities in both time and frequency sampling domains, whereas NUFFT

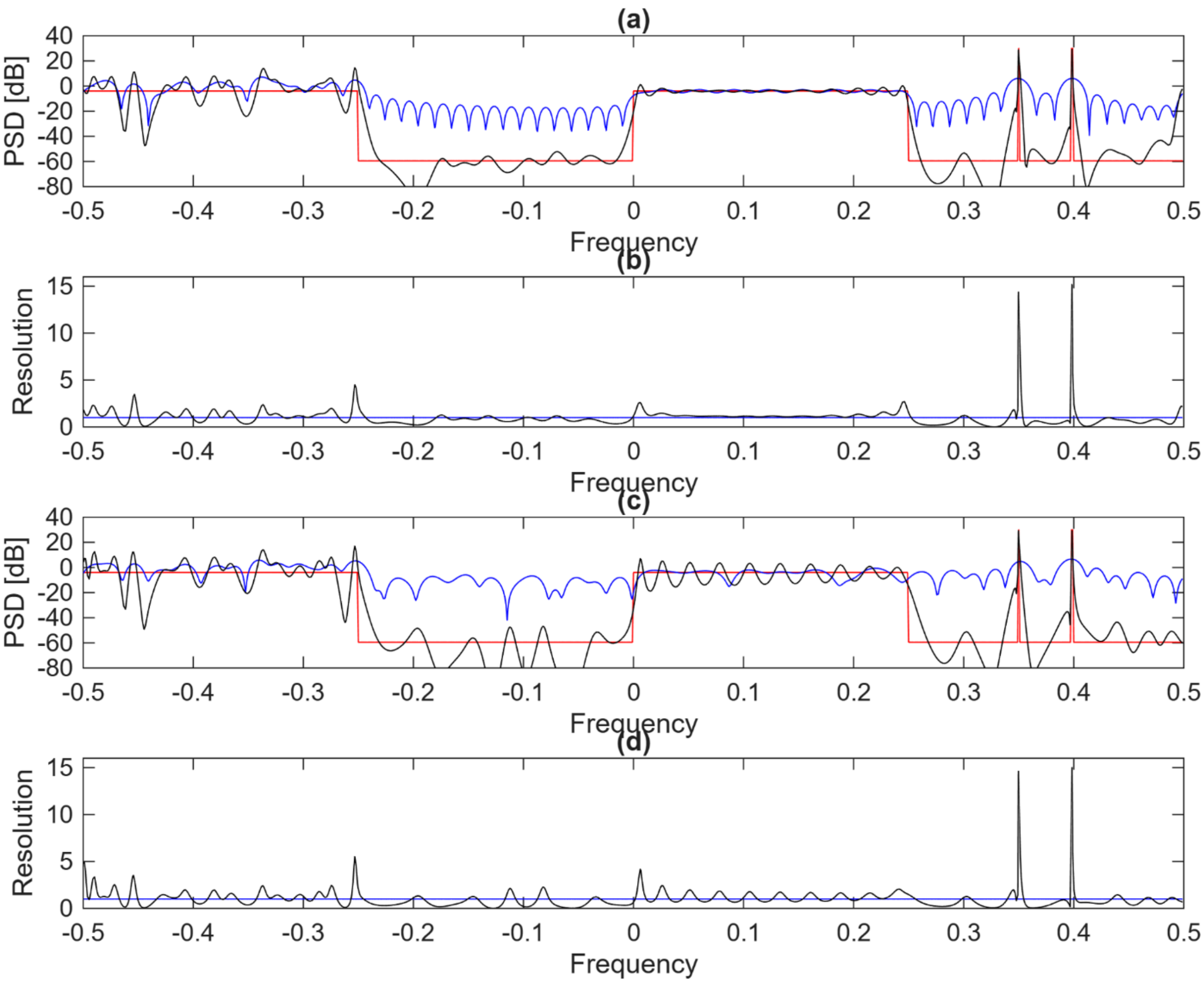

**Figure 3 – EDFT and NUFFT performance on nonuniform frequency grids.**

**(a)** PSD of uniform sequence analyzed on a randomized nonuniform frequency grid;
**(b)** Relative frequency resolution showing grid-dependent shifts on deterministic components;
**(c)** PSD of nonuniform sequence evaluated on the same nonuniform grid;
**(d)** Corresponding relative resolution demonstrating EDFT stability under arbitrary frequency sampling.

performance degrades noticeably. This highlights EDFT's robustness under dual nonuniformity.

Figures 1–3 collectively demonstrate the robustness and universality of EDFT. The method operates reliably on uniform and nonuniform sequences and with uniform or nonuniform frequency grids, delivering high-resolution spectral estimates in all cases. However, when nonuniformities are introduced, the use of fast computational algorithms (FFT, IFFT, Levinson-Durbin recursion) within EDFT becomes less feasible, increasing computational cost. Later simulations highlight cases where such tradeoffs remain justified due to the performance advantages of EDFT.

Figure 4 examines the performance of EDFT when the analysis bandwidth is extended to two Nyquist zones by doubling the spectral range to $[-1, 1)$ Hz and increasing the number of frequency samples to $N = 2000$. This setting allows evaluation of EDFT under wideband analysis conditions and highlights its behavior relative to FFT and NUFFT.

**Fig. 4(a)**—Under uniform sampling, both FFT and EDFT exhibit the expected periodic spectral repetition intrinsic to the classical DFT. The deterministic and random components reappear symmetrically in the second Nyquist zone, reflecting the aliasing structure of uniformly sampled data.

**Fig. 4(b)**—For nonuniformly sampled input, EDFT behaves fundamentally differently: no periodic spectral replicas appear beyond 0.5 Hz (the Nyquist frequency). Instead, EDFT

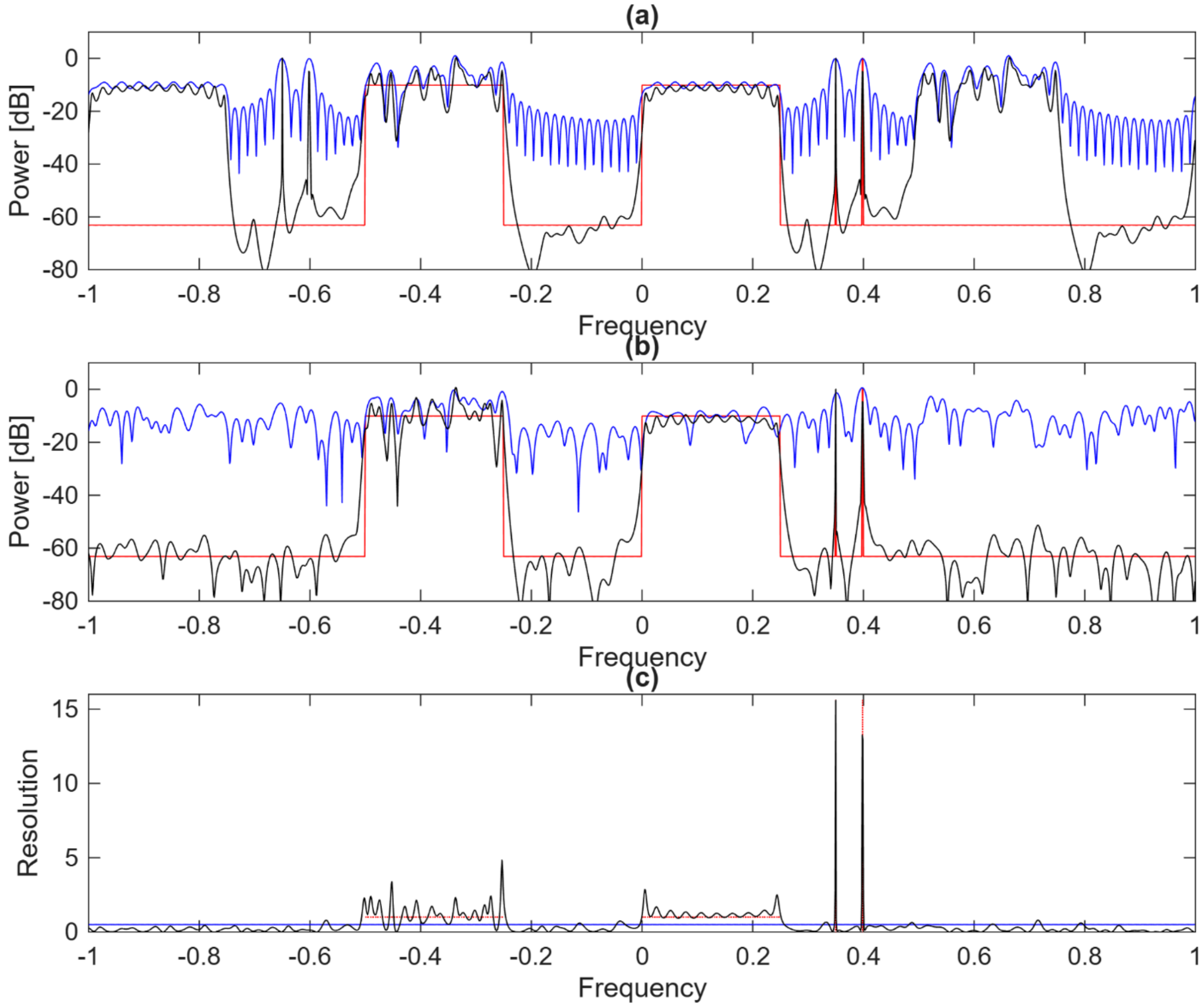

**Figure 4 – EDFT and FFT/NUFFT performance across two Nyquist zones.**

**(a)** Spectral estimates of uniformly sampled sequence using FFT and EDFT over doubled frequency span;
**(b)** NUFFT and EDFT estimate for nonuniform sequence, showing correct localization above 0.5 Hz by EDFT;
**(c)** Relative frequency resolution for NUFFT and EDFT when processing two Nyquist zones.

correctly reports only the background ADC noise floor outside the physical bandwidth of the signal. This demonstrates a key theoretical property—nonuniform sampling removes the strict periodicity constraint of the DFT and prevents artificial spectral replication.

**Fig. 4(c)**—The relative resolution curves reveal that NUFFT resolution is reduced by half (≈0.5) due to processing over two Nyquist zones. In contrast, EDFT maintains its adaptive resolution behavior, achieving peaks of approximately $N/K \approx 2000/64 \approx 31$, near the deterministic components, preserving standard resolution (~1) on random and transient components, and decreasing it in noise-only regions. This illustrates EDFT's ability to redistribute resolution while respecting the *global resolution constraint* and maintaining support for a single Nyquist zone (the squares under the blue and black plots are the same as in Fig. 2d).

Figure 5 discusses the impact of missing samples on spectral estimation and compares the behavior of NUFFT with EDFT under increasingly sparse time-domain data. Two degraded sequences are created by randomly removing 16 and 24 samples from the original 64-point uniformly sampled signal, increasing the mean sampling period to $T_s = 64T/48 = 1.33$ s and $T_s = 64T/40 = 1.60$ s, respectively.

**Fig. 5(a)**—The baseline spectrum of the complete 64-sample sequence (identical to Fig. 1(b)) is shown for reference. This serves as the ground truth benchmark for evaluating the effects of sample removal.

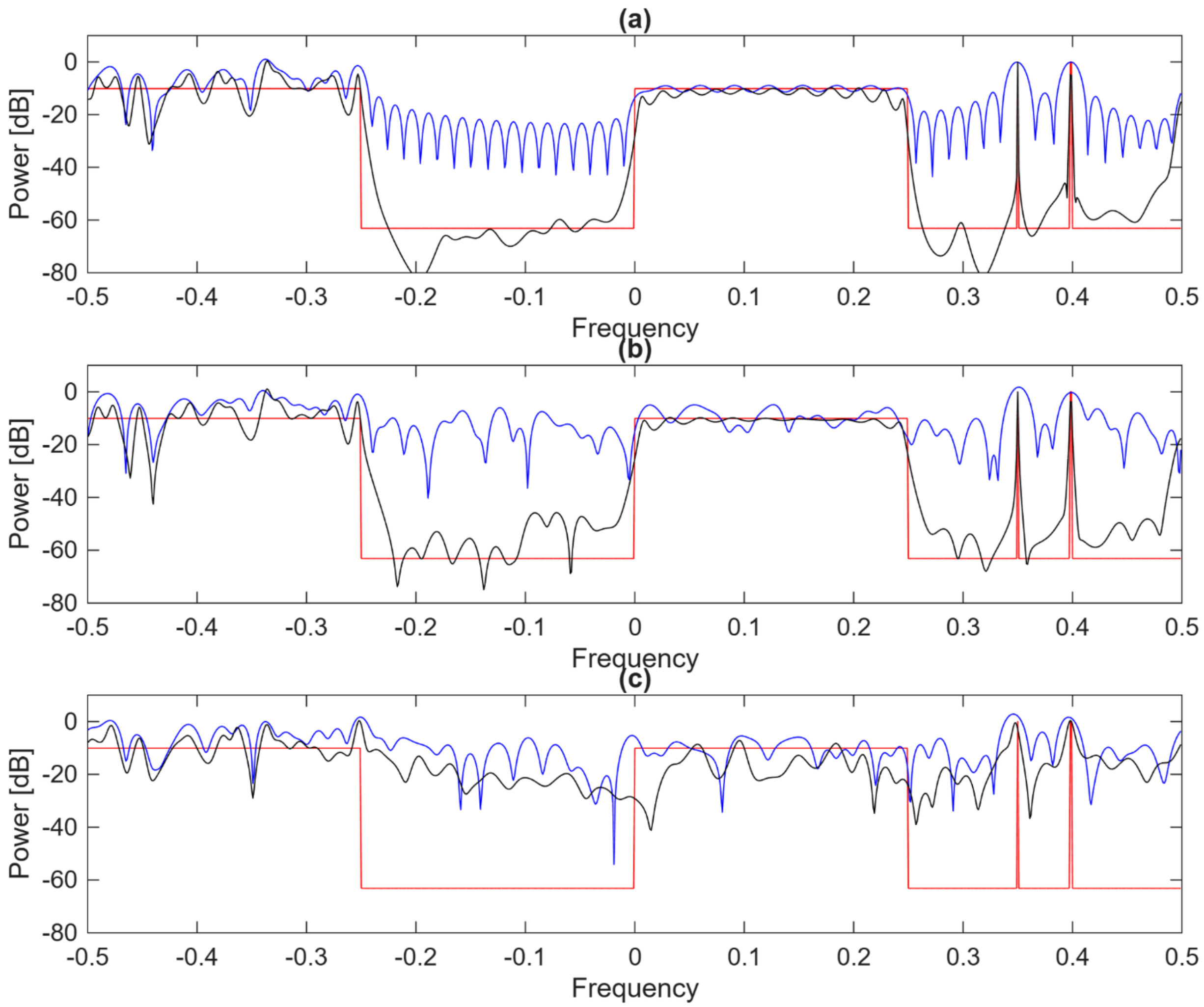

**Figure 5 – Effect of missing samples on NUFFT and EDFT.**

**(a)** Power spectrum of the complete 64-sample sequence;
**(b)** Spectrum with 16 randomly removed samples, illustrating NUFFT breakdown and EDFT robustness when spectral support remains within one Nyquist zone;
**(c)** Spectrum with 24 missing samples showing degradation of EDFT when the one-zone constraint is violated.

**Fig. 5(b)**—With 16 samples removed, NUFFT becomes unstable: the missing data introduce irregular gaps that lead to severe spectral leakage and corrupted PSD estimates. In contrast, EDFT maintains accurate spectral reconstruction as long as the total occupied bandwidth does not exceed one Nyquist zone. This demonstrates EDFT's resilience to moderately incomplete or irregular sampling patterns.

**Fig. 5(c)**—When 24 samples are removed, the mean sampling rate becomes too low to support the full spectral content of the signal. Both EDFT and NUFFT fail in this regime. EDFT performance deteriorates once the single-zone spectral support condition is violated, confirming the theoretical constraint derived earlier. This illustrates the fundamental limit: EDFT can compensate for missing samples only when the effective sampling density remains sufficient to capture the signal's bandwidth.

Sensitivity to sample removal varies by component type. Deterministic sinusoids (pure tones) are the most tolerant to missing samples, especially when aligned with the EDFT frequency grid. Transient components, however, require dense local sampling and degrade rapidly when gaps occur, reflecting their inherently broadband nature.

## B. EDFT Performance on Deterministic Signals with Sparse Sampling

This simulation evaluates the capability of EDFT to process deterministic signals when the average sampling interval $T_s$ is increased far beyond the nominal sampling period $T$, as anticipated in **[8]**. In this experiment, the signal consists solely of four sinusoids embedded in additive white Gaussian noise, and the sampling pattern is made deliberately sparse and highly irregular—representing an extreme undersampling scenario.

A real-valued sequence of length $K$ = 64 is synthesized as the sum of four sinusoids with amplitudes 0.5, 1, 2, and 3, each with an arbitrary phase and a frequency drawn randomly from the EDFT grid with spacing 0.001 Hz. The analysis grid spans $NT$ =1000 s, and the sampling instants are chosen by selecting 64 points uniformly at random within this interval. This produces an approximately sixteenfold increase in the mean sampling period, $T_s = NT/K$ = 1000/64 = 15.625 s, while retaining only 64 actual measurements. White Gaussian noise with SNR = 20 dB is added. The resulting sequence is shown in **Fig. 6(a)**.

**Fig. 6(b)** compares the true sinusoidal amplitudes (red markers) with the corresponding NUFFT (blue) and EDFT (black) spectral estimates. Only the nonnegative frequencies (500 points) are shown because the signal is real-valued, and therefore possesses conjugate spectral symmetry.

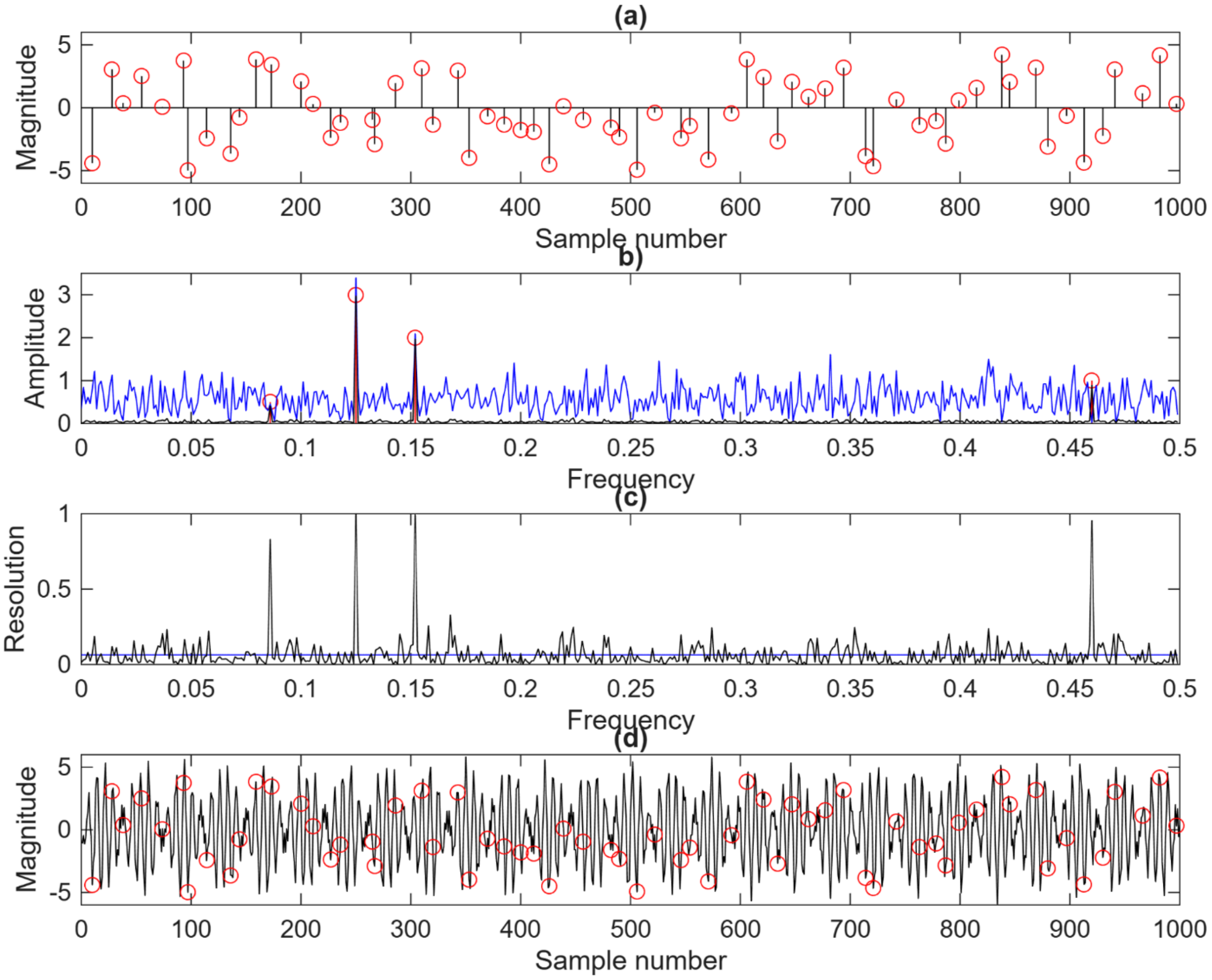

**Figure 6 – EDFT performance on sparse deterministic sequences.**

**(a)** Real-valued 64-point sequence sampled at randomly selected time points over a 1000-s interval;
**(b)** True sinusoidal amplitudes versus NUFFT (blue) and EDFT (black) estimates;
**(c)** Relative frequency resolutions showing EDFT's ability to achieve near-ideal resolution despite 16× undersampling;
**(d)** Reconstructed and interpolated time-domain sequence from EDFT illustrating accurate recovery of deterministic components.

The NUFFT fails to detect the lower-power components and exhibits substantial amplitude bias at the stronger frequencies. In contrast, EDFT accurately recovers all four sinusoids—both their frequencies and amplitudes—despite the severe sparseness of the observations.

This performance difference is explained by the relative frequency resolutions shown in **Fig. 6(c)**. For the NUFFT operating with mean sampling period $T_s$, the relative resolution equals $1/(2f_uT_s) = K/N = 0.064$, well below unity and therefore insufficient for reliable spectral estimation. Because the time span covers nearly 16 Nyquist zones, the NUFFT suffers from aliasing and leakage. EDFT avoids this limitation. Its relative resolution, $\frac{1}{2f_uT_sK}\mathbf{F./S} = \frac{1}{N}\mathbf{F./S}$, is amplified by a factor of $N/K$, enabling EDFT to reach near-*normal resolution* (≈1) at the components even when severe undersampling renders the NUFFT fundamentally unusable.

**Fig. 6(d)** shows the time-domain reconstruction obtained by applying the IFFT to the EDFT spectrum. The reconstructed sequence of length $N = 1000$ consists of the original 64 samples plus 936 interpolated values generated by EDFT. Only the deterministic portion of the signal is interpolated; the added white noise remains highly localized around the actual sampling instants, shown as red circles.

Collectively, Figure 6 demonstrates that EDFT can accurately analyze, resolve, and reconstruct deterministic signals even under extreme undersampling conditions—well beyond the operational limits of classical DFT-based methods.

## C. Analysis of the Marple & Kay Data Set and Extrapolation Behavior

The following simulations employ the classical Marple & Kay data set **[3]**, a 64-point real-valued sequence composed of three sinusoids and colored noise component. Two of the sinusoids have unit power at frequencies 0.20 Hz and 0.21 Hz, forming a closely spaced pair that is notoriously difficult for conventional spectral estimators to resolve. A third sinusoid, with amplitude 0.1 (20 dB lower), is located at 0.10 Hz. In addition, colored noise occupies the frequency interval [0.2, 0.5] Hz. The resulting "true" spectrum—comprising both discrete and continuous components—is shown as red curves in Figure 7. The upper frequency is $f_u = 0.5$ Hz, and the DFT length is set to $N = 1000$, yielding the same fine frequency grid used in previous simulations.

Figure 7 presents a side-by-side comparison of FFT, EDFT, and HRDFT spectral estimates for this benchmark dataset. Although these spectra have been shown individually in earlier works (e.g., **[6]**, **[10]**), their direct juxtaposition here highlights the fundamental differences among the methods. Additional comparisons with other classical spectral estimation techniques (including the Capon/MVDR method) can be found **[3]**. The plots reveal the following:

**Fig. 7(a)**—The FFT power spectrum. Due to its limited frequency resolution, the FFT cannot separate the two adjacent sinusoids at 0.20 Hz and 0.21 Hz, which merge into a single broadened peak. Nevertheless, the FFT adequately reflects the overall shape of the colored-noise background.

**Fig. 7(b)**—The EDFT spectrum after 15 iterations. The first iteration coincides with the FFT, but subsequent iterations selectively increase resolution around strong spectral components while reducing resolution in low-power regions. As a result, all three sinusoids are distinctly resolved—including the challenging 0.20/0.21 Hz pair—and the continuous colored-noise spectrum is faithfully reconstructed.

**Fig. 7(c)**—The HRDFT spectrum after 15 iterations. Consistent with observations in **[10]**, HRDFT is well-suited for line spectral estimation: it successfully resolves all three sinusoids, including the low-amplitude 0.10 Hz component. However, HRDFT collapses the broadband colored-noise region into a small number of narrow peaks, creating a misleading representation

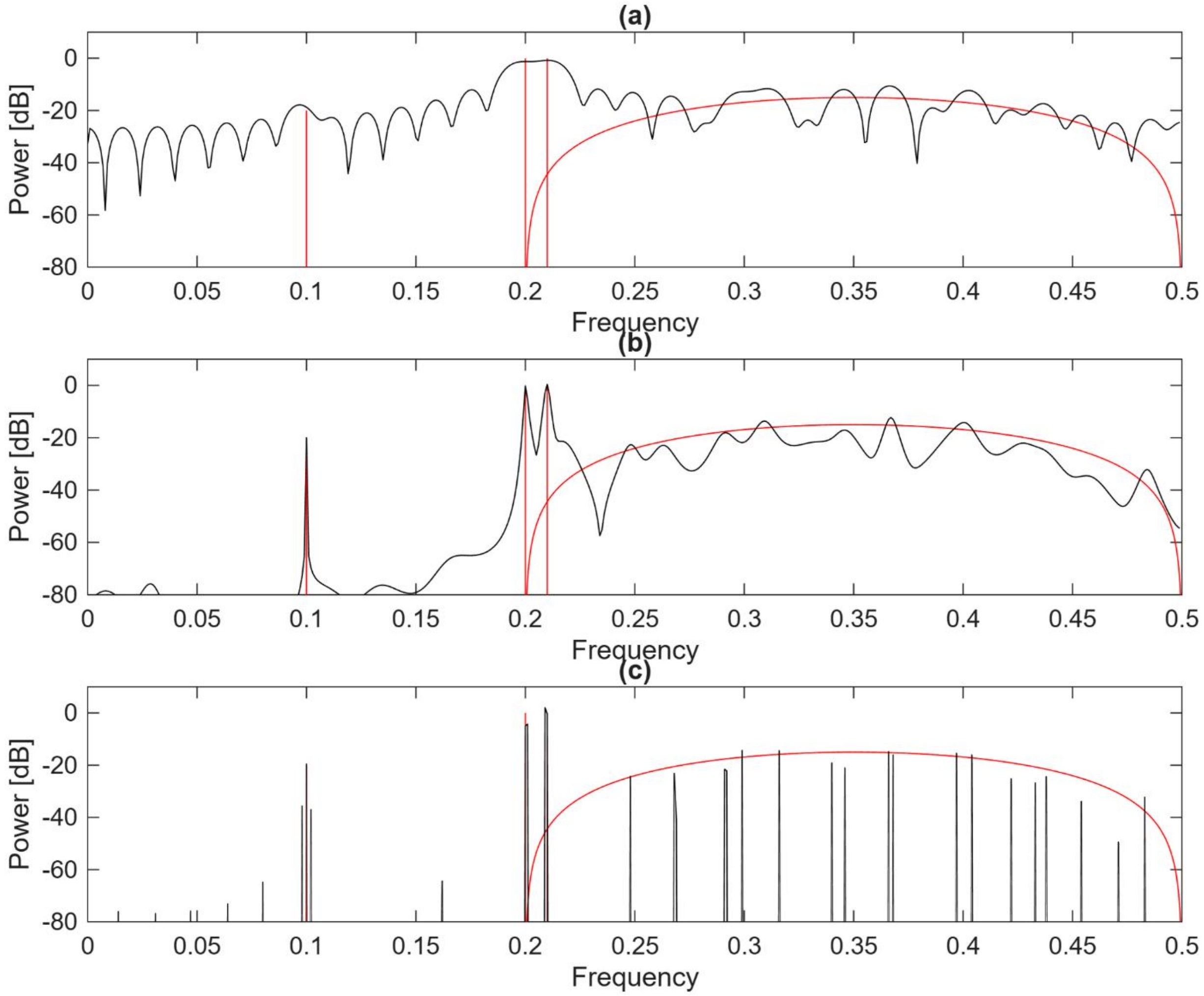

**Figure 7 – Marple & Kay data set: spectral estimates.**

**(a)** FFT power spectrum showing insufficient resolution for closely spaced sinusoids;
**(b)** EDFT power spectrum resolving all sinusoidal components and accurately capturing colored-noise continuum;
**(c)** HRDFT power spectrum emphasizing line components but failing to represent broadband structure.

of the continuous spectrum. Thus, HRDFT is advantageous for purely deterministic signals but inadequate for signals containing both discrete and broadband components.

Overall, Figure 7 demonstrates that while FFT and EDFT can reconstruct both discrete and continuous spectral components, EDFT provides superior resolution and accuracy, whereas HRDFT performs well only when the spectrum is dominated by sinusoids.

A distinctive property common to all three transforms—FFT, EDFT, and HRDFT—is their ability to reproduce the original input sequence exactly when the IFFT is applied to their respective spectral outputs. Because the transform length is $N$ = 1000 while the Marple & Kay data set contains only 64 samples, the inverse transform returns 936 additional samples, forming forward and backward extrapolations of the known data. In Figure 8 samples 65, 66, 67, … correspond to forward extrapolation, samples 1000, 999, 998, … correspond to backward extrapolation, and the original 64 samples are shown in blue.

The extrapolated sequences behave markedly differently for the three transforms:

**Fig. 8(a)**—FFT produces zero-valued extrapolation outside the data window. This result stems from the implicit periodic extension with zero-padding—effectively assuming that all unknown samples are identically zero. The consequence is a flat, uninformative extrapolation that reflects the FFT's rigid boundary assumptions.

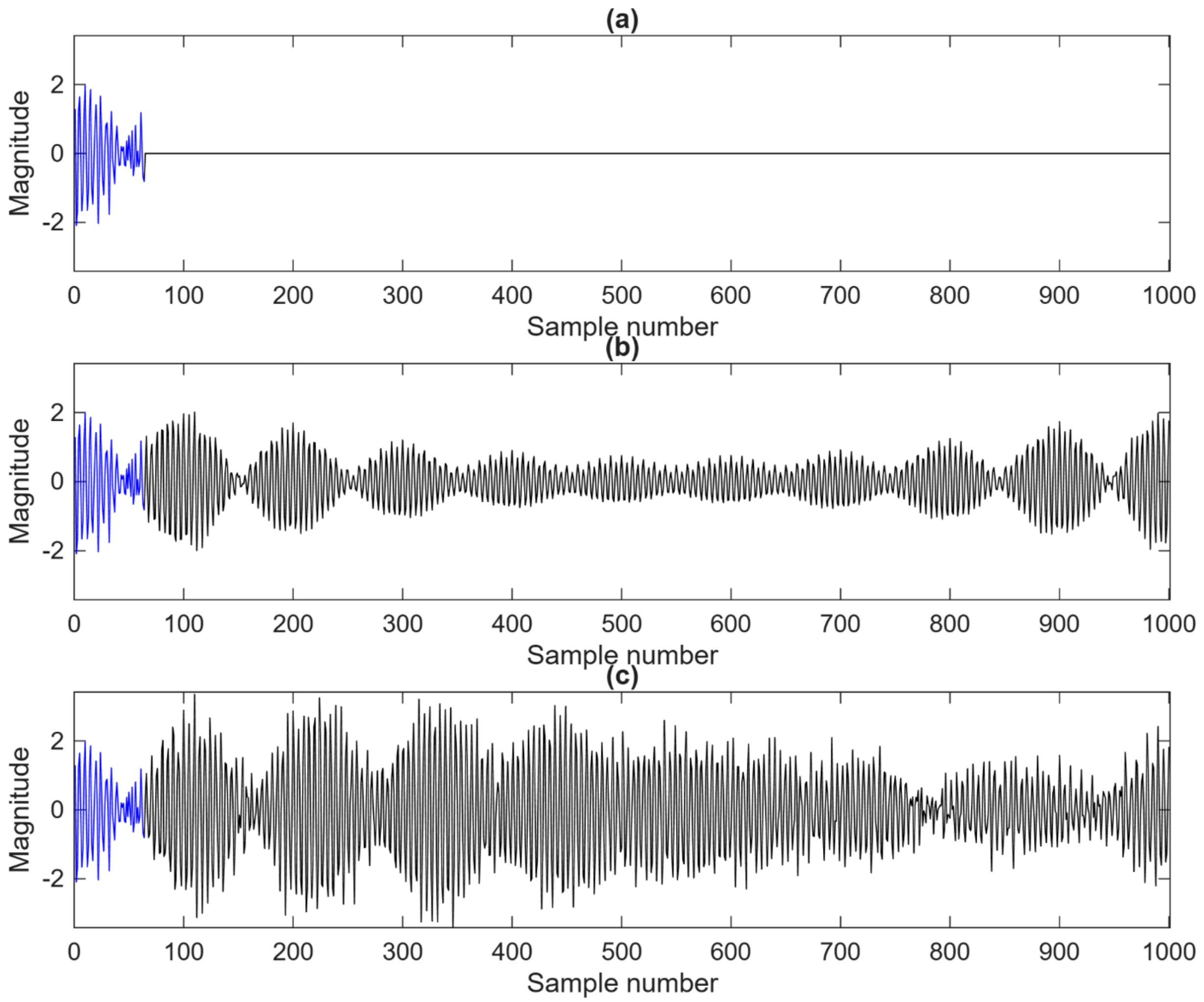

**Figure 8 – Time-domain extrapolation of Marple & Kay sequence.**

Forward and backward extrapolations obtained by applying the inverse transform to the outputs of **(a)** FFT, **(b)** EDFT, and **(c)** HRDFT. EDFT extrapolation exhibits physically plausible continuation, FFT produces zero-padding behavior, and HRDFT yields unrealistically increasing oscillations.

**Fig. 8(b)**—EDFT predicts a continuation of the signal beyond the observed interval that retains a magnitude similar to the known data but gradually decreases in amplitude over time. This behavior is consistent with EDFT's adaptive weighting and resolution redistribution, avoiding artificial discontinuities while not enforcing strict periodicity.

**Fig. 8(c)**—HRDFT reconstruction exhibits oscillations that grow in amplitude outside the data window, implying an unrealistic increase in signal energy. This behavior arises because HRDFT converges to a sparse line-spectrum model that cannot accommodate the broadband noise component of the Marple & Kay data, leading to an extrapolation dominated by exaggerated sinusoidal components.

These extrapolation patterns reflect the fundamental modelling assumptions of each transform—FFT implicitly assumes periodic extension with zeros beyond the window; HRDFT assumes a small number of sinusoids; EDFT provides a balanced, data-adaptive estimate that remains statistically consistent with the observed sequence.

To further investigate the extrapolation properties observed in Figure 8, the Marple & Kay sequence is replaced by a 64-point realization of white Gaussian noise. In theory, the power spectral density (PSD) of white noise is flat and its samples are uncorrelated, implying that no meaningful extrapolation is possible—values outside the observed interval should exhibit no predictable structure.

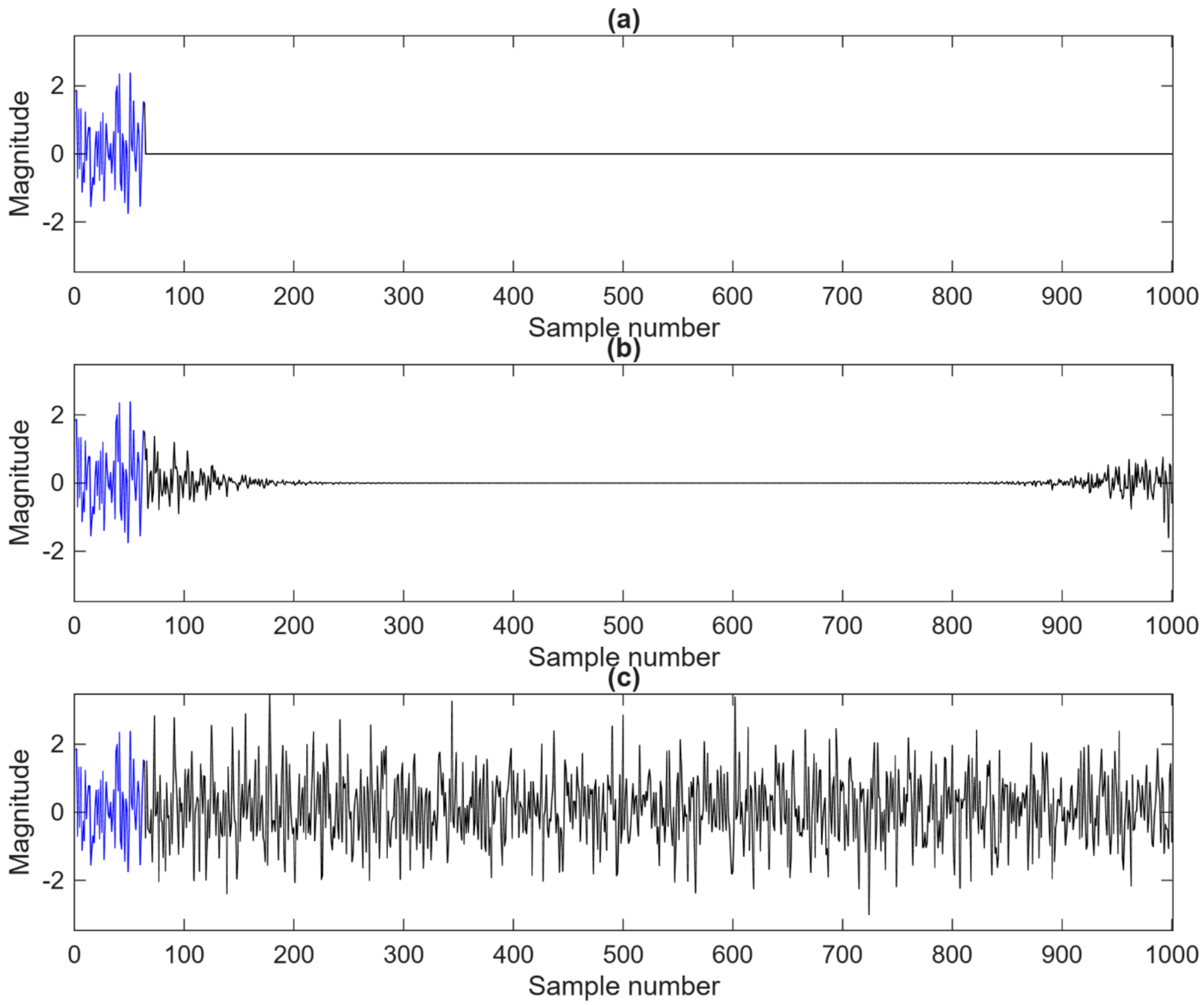

**Figure 9 – Extrapolation of white Gaussian noise.**

**(a)** FFT preserves theoretical behavior by producing zeros beyond the finite observation interval;
**(b)** EDFT yields rapidly decaying extrapolated values consistent with finite-length noise realizations;
**(c)** HRDFT incorrectly generates long-term correlated extrapolation.

**Fig. 9(a)**—As in the previous experiment, the FFT reconstruction yields exactly zero extrapolation outside the 64 known samples. For white noise, this behavior is fully consistent with theory: since the process is memoryless, no information is available for forecasting or backcasting.

**Fig. 9(b)**—The EDFT extrapolation rapidly decays to zero as time moves away from the observed interval. This behavior is theoretically sound. When processing broadband noise confined in a single Nyquist zone, EDFT assigns nearly uniform resolution across frequencies, leading to a smooth decay that reflects the finite duration of the observed noise realization without introducing artificial structure.

**Fig. 9(c)**—In contrast, HRDFT produces a long, coherent extension over all 1000 samples, giving the false impression of strong temporal correlation and oscillatory patterns. This is inconsistent with the properties of white noise and highlights HRDFT's tendency to impose a sparse line-spectrum model even when the underlying process is entirely stochastic.

Taken together, Figures 8 and 9 clearly demonstrate that EDFT is the only method that produces physically meaningful extrapolations for both deterministic and stochastic signals. FFT fails to extrapolate, while HRDFT generates unrealistic, highly structured extensions when broadband components are present.

## D. Processing of a Gaussian Pulse by EDFT, FFT, and NUFFT

The final simulation evaluates the ability of the proposed EDFT algorithm to process a Gaussian-modulated sine wave—a signal for which the Fourier transform is known analytically, enabling a direct and precise assessment of spectral accuracy. The Fourier transform of the Gaussian-modulated cosine is

$$e^{-at^2}\cos(2\pi f_c) \overset{FT}{\leftrightarrow} \sqrt{\frac{\pi}{4a}}\left(e^{-\frac{\pi^2(f-f_c)^2}{a}} + e^{-\frac{\pi^2(f+f_c)^2}{a}}\right), \quad (61)$$

with parameter $a = \frac{(\pi f_c B_W)^2}{4\log 10^{-0.3}}$, central frequency $f_c = 0.15$ Hz and fractional bandwidth

$$B_W = \frac{(f_{max}-f_{min})}{f_c} = \frac{(1.1f_c - 0.9f_c)}{f_c} = 0.2.$$

Uniform and nonuniform sequences of length $K = 64$ are generated directly from (61) using sampling period $T = T_s = 1$ s. To make the problem nontrivial, the Gaussian pulse is centered at $t = 61$ s so that only a part of the pulse lies within the available data window (Figure 10b, blue).

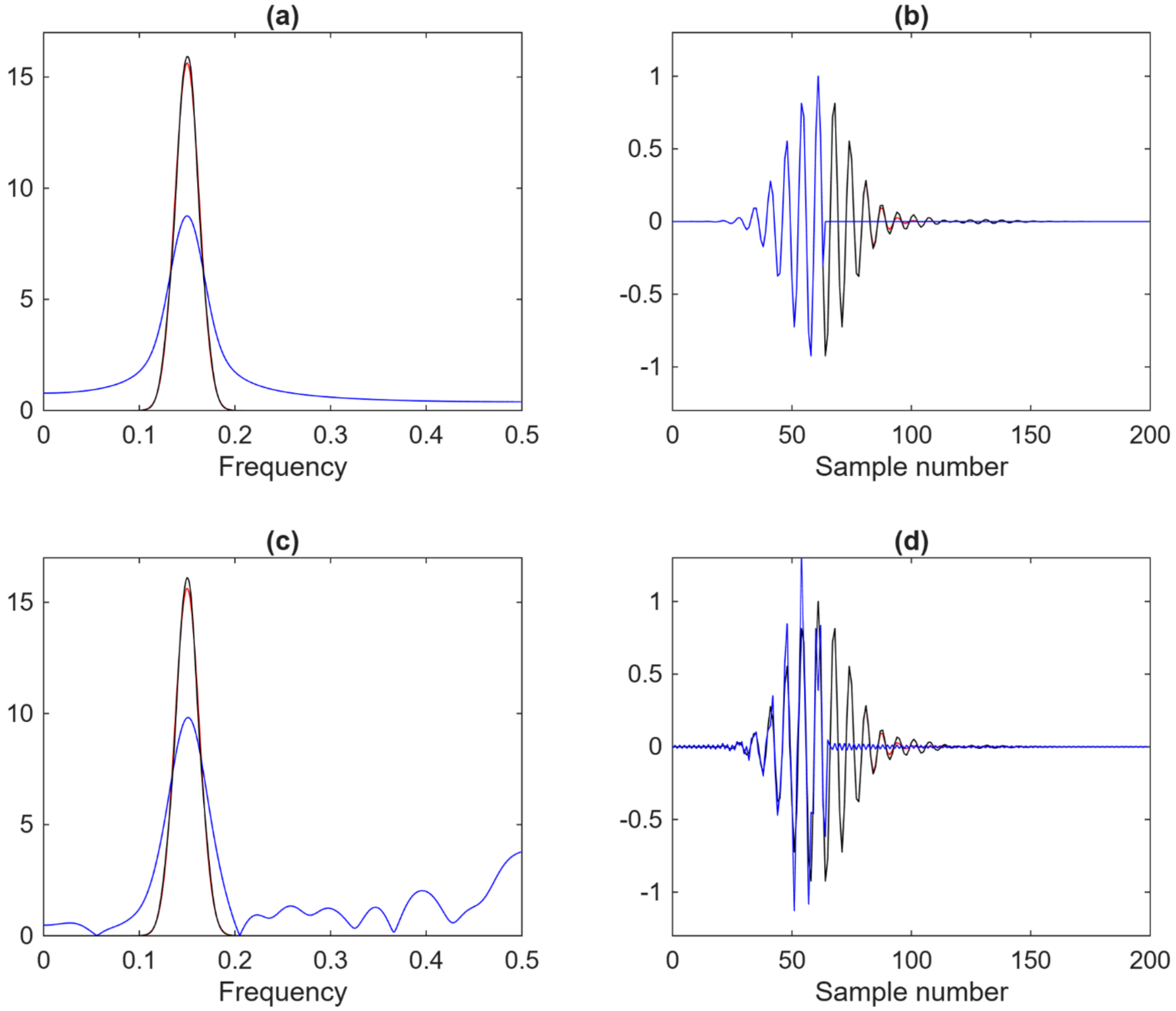

**Figure 10 – EDFT and FFT/NUFFT applied to Gaussian-modulated sinusoid.**

**(a)** EDFT for uniform input showing near-perfect match with analytic Fourier transform versus FFT;
**(b)** Reconstructed pulse demonstrating EDFT's ability to recover the full waveform, while FFT returns only observed segment;
**(c)** EDFT for nonuniform input maintaining accuracy, unlike NUFFT;
**(d)** Reconstructed pulse showing EDFT's distortion-free recovery, whereas NUFFT introduces high-frequency artifacts.

The transform length is set to $N = 1000$, and frequencies lie on the FFT grid.

Figures 10(a) and 10(c) compare EDFT, FFT, and NUFFT spectra with the analytic Fourier transform. For both uniform and nonuniform sampling:

- EDFT (black) matches the analytic spectrum (red) with near-perfect accuracy.
- FFT (blue) exhibits pulse width and high deviations.
- NUFFT (blue) amplifies high-frequency components under nonuniform sampling, producing artificial broadening inconsistent with the theoretical Gaussian shape.

The superior EDFT performance arises from its adaptive weighting and resolution redistribution, which naturally accommodate localized and smooth spectral profiles—properties that classical DFT-based methods do not exploit.

These spectral differences manifest clearly in the time domain. Figures 10(b) and 10(d) show the reconstructed pulse obtained via inverse transforms:

- FFT returns only the observed portion of the pulse, zero-padding the remainder.
- NUFFT reconstruction deviates strongly from the true waveform, especially in amplitude and tail behavior.
- EDFT reconstruction preserves the pulse shape, provides a smooth and physically plausible extrapolation beyond the observed interval, and converges in few iterations.

The modest iteration count is expected: numerical accuracy is limited by the fact that the input is a mathematically ideal function rather than a physical measurement, making the solution more sensitive to finite-precision arithmetic. Despite this, EDFT achieves excellent agreement with the analytic Fourier transform, confirming its robustness for both uniform and nonuniform data.

## E. EDFT and its Inverse (IEDFT) in Time & Frequency Domains

Figure 11 summarizes the four principal configurations in which the EDFT and its inverse, the IEDFT, may be applied. These configurations arise from the combinations of uniform and nonuniform sampling in both the time and frequency domains:

1. Uniform input → Uniform output (standard case)
2. Nonuniform input → Uniform output (Type 1)
3. Uniform input → Nonuniform output (Type 2)
4. Nonuniform input → Nonuniform output (Type 3)

These relationships are represented by circular diagrams—**Time** and **Frequency**—each containing $N$ query points corresponding to the transform length. Filled dots indicate the positions of the $K$ available samples ($K \leq N$); unfilled locations represent values reconstructed by the IEDFT (interpolation or extrapolation). Bidirectional arrows illustrate the four possible EDFT/IEDFT mappings, highlighting the method's ability to accommodate arbitrary nonuniformity in either domain.

### Time Domain Representation

On the **Uniform Time** circle, adjacent points are equally spaced by the sampling period $T$.

On the **Nonuniform Time** circle, these spacings vary, and only the mean sampling period $T_s$ can be defined. In Figure 11, the nonuniform circle corresponds to processing within a single Nyquist zone, so $T_s = T$. Filled dots indicate the actual measurement times; unfilled positions are reconstructed by the IEDFT—either as interpolated values or extrapolated values when $N > K$.

### Frequency Domain Representation

On the **Uniform Frequencies** circle, the frequency spacing is $\Delta f = \frac{2f_u}{N}$, covering the range $[-f_u, f_u)$ and placing the frequency $-f_u$ diametrically opposite the zero frequency $f_0$.

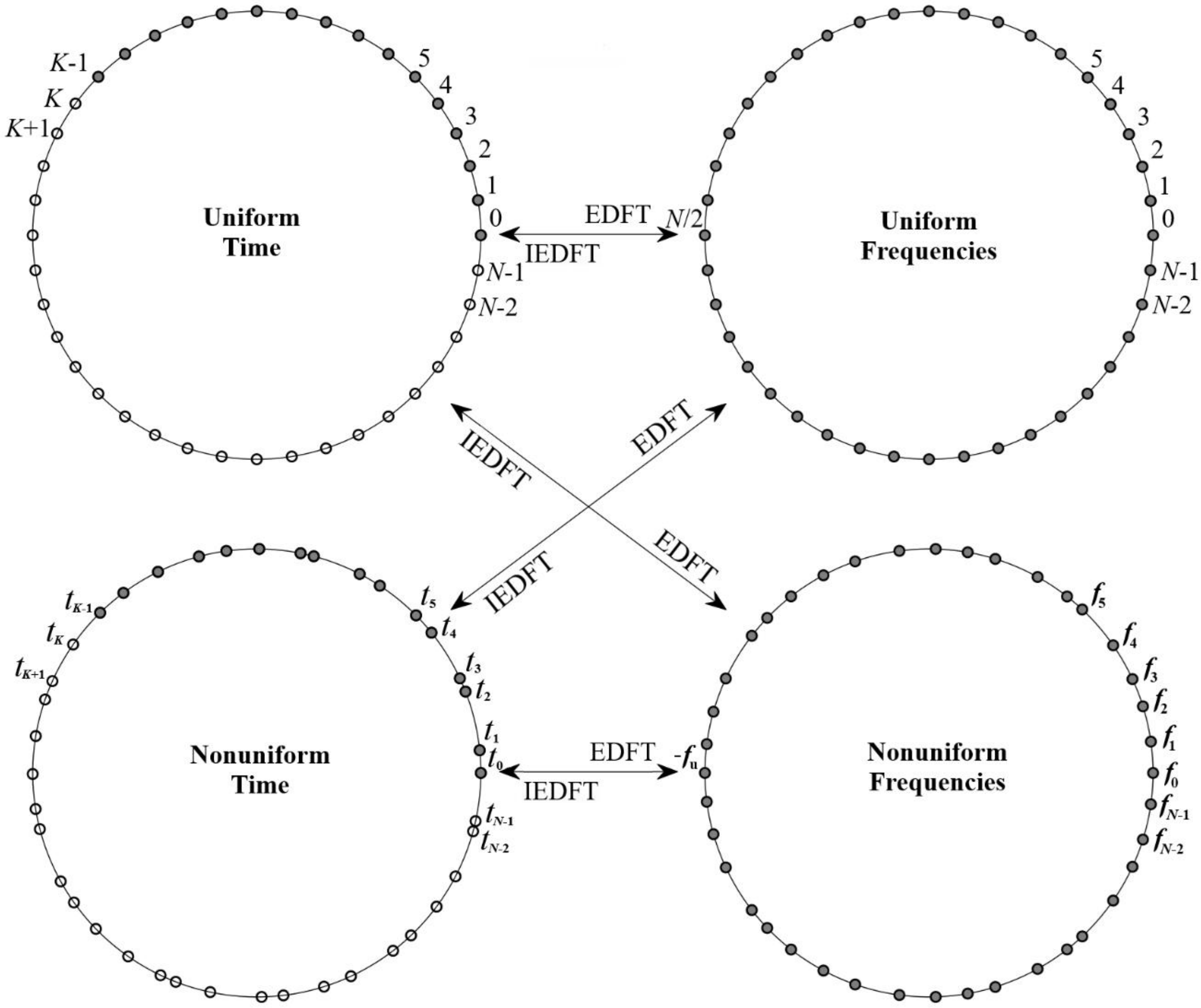

**Figure 11 – Conceptual interpretation of EDFT/IEDFT mappings.**

Diagram illustrating four EDFT applications involving uniform and nonuniform time and frequency grids. Closed circles represent query points of length *N*; filled dots mark available samples. Bidirectional arrows depict forward and inverse transforms across uniform/nonuniform domains, emphasizing EDFT's ability to accommodate arbitrary nonuniformity in either domain.

If $N$ is even, indices span [0, $N$-1], with $-f_u$ at $N/2$ (as in Figure 11).

If $N$ is odd, $-f_u$ is placed midway between two grid points—with indices $(N-1)/2$ and $(N+1)/2$.

On the **Nonuniform Frequencies** circle, the spacing is arbitrary, covering the range $[-f_u, f_u)$ and keeping $-f_u$ opposite to $f_0$. For real-valued inputs, every positive frequency must have a symmetric negative counterpart (e.g., $f_1 = -f_{N-1}$, $f_2 = -f_{N-2}$, …), a requirement arising from Euler's identity: $e^{ix} = \cos(x) + i\,sin(x)$, which ensures that the IEDFT produces real-valued samples.

## Nyquist-Zone Considerations

The product $2f_uT_s$ determines the number of Nyquist zones in which a nonuniform sequence is effectively sampled. If the EDFT covers more than one zone, the IEDFT must be evaluated on a time grid dense enough to satisfy the Nyquist criterion for the reconstructed signal. In such cases, the transform length must satisfy $N \geq 2f_uT_sK$. This condition guarantees that the recovered time-domain sequence is represented without aliasing.

### Relation to Simulation Results

The configurations in Figure 11 correspond directly to the simulations in Sections VI.A–D:

**Figures 1 to 3**—Single Nyquist zone processing in Uniform Time / Uniform Frequencies domains (Fig. 1), Type 1 (Nonuniform input → Uniform output) in Fig. 2, and Types 2 / Type 3 processing on nonuniform frequency grids.

**Fig. 4**—Two Nyquist zones; Time circles require interleaved unfilled dots and the IEDFT uses a grid twice as dense.

**Fig. 5 and 6**—Sequences with missing samples in 1.33, 1.6 and ~16 Nyquist zones; Correspond to special cases on the Uniform Time circle where filled dots are removed, creating nonuniform sampling.

In all scenarios, EDFT provides the forward spectral estimate, and the IEDFT (or IFFT when the frequencies lie on a uniform FFT grid) supplies the reconstructed values at unfilled time points. These reconstructions—whether interpolated or extrapolated—remain consistent with the structure of the underlying signal and the spectral model.

## VII. EDFT MATLAB and OCTAVE Packages

A software package implementing the Extended Discrete Fourier Transform (EDFT) has been developed in MATLAB and OCTAVE to support the theoretical and algorithmic results presented in this paper. The package provides accessible reference implementations of the EDFT, enabling users to reproduce the simulations, study convergence behavior, and apply the method to their own uniform or nonuniform data sets. Each function includes built-in help text describing syntax, algorithmic details, usage and features.

The package consists of five programs:

**edft.m**—The core implementation of the EDFT algorithm for uniform and nonuniform sequences and frequency grids. Three algorithmic modes are provided:

- **Algorithm 1**—fast solver for uniformly sampled data using MATLAB library functions FFT, IFFT, and Levinson-Durbin recursion to accelerate computations;
- **Algorithm 2**—variant supporting missing samples encoded as NaN (Not a Number);
- **Algorithm 3**—general solver for fully nonuniform time and/or frequency sampling.

The inverse transform for Algorithms 1 and 2 is MATLAB's built-in IFFT.

**iedft.m**—Computes the IEDFT on either uniform or nonuniform time grids. This function uses MATLAB library functions IFFT or the adjoint NUFFT to accelerate calculations, supporting interpolation and extrapolation of sequences.

**edft2.m**—A two-dimensional extension of the EDFT suitable for image and multidimensional signal processing. The program mirrors the structure of MATLAB's FFT2, replacing internal FFT calls with EDFT computations. The corresponding inverse transform is performed using MATLAB's IFFT2. This enables high-resolution spectral analysis in 2-D domains.

**demoedft.m**—A demonstration script that evaluates EDFT performance over successive iterations for the simulated test signal. It also provides a comparison with MATLAB's NUFFT implementation, allowing users to explore accuracy differences between EDFT and conventional transforms.

**edft_fig.m**—A script that reproduces all figures presented in this paper, providing a complete template for numerical experimentation and result verification.

The original version of the EDFT software (under the name **gdft.m**) was released on MATLAB File Exchange on October 7, 1997 as MATLAB 4.1 code. A substantially updated MATLAB implementation was uploaded on August 5, 2006 and is available through mathworks.com and researchgate.net.

The OCTAVE version of the package was released on July 4, 2024 and is available on researchgate.net, providing full open-source accessibility for non-MATLAB users.

# VIII. Conclusion

This paper presented a comprehensive theoretical formulation and practical implementation of the Extended Discrete Fourier Transform (EDFT). Beginning with the Fourier integral and its orthogonality properties, we derived EDFT as the solution to weighted least-squares estimators in both continuous and discrete domains. Unlike heuristic modifications of the DFT, EDFT emerges as a mathematically rigorous generalization that adapts the Fourier basis to the power distribution of the signal, thereby preserving linearity, invertibility, and full compatibility with classical Fourier analysis while substantially improving spectral resolution.

The EDFT framework provides two properly scaled outputs: one proportional to the power spectral density and another to the power of the signal components. Their ratio defines the local frequency resolution, enabling EDFT to adaptively redistribute resolution across the spectrum, and at the same time serve as a detector of a sinusoidal signal. We demonstrated that maximum achievable resolution depends on the transform length rather than the number of observed samples—contrary to a common misconception rooted in classical DFT interpretation. Thus, the inverse EDFT not only reconstructs input data without distortion but also extrapolates and interpolates the sequence up to the full transform span.

Any method claiming high-resolution spectral capability necessarily embeds assumptions about the signal outside the observation window, consistent with the uncertainty principle. The advantage of EDFT is that its enhanced resolution arises from the minimum-error weighted least-squares solution, ensuring stability, reliability, and physical consistency.

A rigorous comparison with established nonparametric approaches—Capon filtering, GWLS, and HRDFT—highlighted fundamental differences. Capon and GWLS require explicit partitioning into signal and noise components, while HRDFT is optimized only for line spectra. EDFT, by contrast, requires no prior decomposition and simultaneously estimates the full complex Fourier spectrum and the adaptive correlation matrix in a manner consistent with the $L^2$-optimal Fourier representation. As a result, EDFT can process deterministic, transient, and stochastic components simultaneously and accurately.

Extensive MATLAB simulations confirmed these theoretical properties. EDFT achieved superior performance for both uniform and nonuniform sampling, suppressed spectral leakage and masking, resolved closely spaced sinusoids, handled incomplete or irregularly sampled data, and produced physically plausible interpolation and extrapolation. The resolution-sum constraint—equal to that of the DFT—was satisfied in all cases, ensuring that resolution enhancements were balanced by reduced resolution elsewhere.

Simulations in one and multiple Nyquist zones further demonstrated EDFT's robustness under highly nonuniform sampling conditions, including cases where the mean sampling period exceeded the nominal sampling interval by more than an order of magnitude. The primary validity condition—that the total spectral support remain within a single Nyquist zone—is both physically intuitive and straightforward to verify.

Finally, EDFT was shown to provide a unified computational framework for the four principal mapping configurations between uniform and nonuniform time- and frequency-domain grids. This flexibility, combined with strong theoretical grounding and practical effectiveness, makes EDFT well suited for modern signal-processing challenges involving irregular sampling, incomplete data, or high-resolution spectral requirements. The method therefore offers a powerful, general-purpose tool that extends the applicability and performance of Fourier-based analysis far beyond the constraints of the classical DFT.

## X. Related Articles

Many thanks to the authors of **[11]** – **[58]** and all others not listed here for using the EDFT approach in their research and applications!

# APPENDIX A – Proof of (17.1)

This appendix provides a rigorous derivation showing that enforcing the target relation containing the Dirac delta distribution $\delta(\omega - \omega_0)$ in the weak sense (or via projections onto the Fourier basis) to integral least-squares estimator (14.1),

$$\Delta = \int_{-\Omega}^{\Omega} \left| 2\pi\delta(\omega - \omega_0) - \int_{-\Theta/2}^{\Theta/2} e^{i\omega_0 t} \alpha(\omega, t) dt \right|^2 d\omega_0,$$

leads directly to the Fredholm-type linear integral equation (17.1) with a *Sinc* kernel:

$$\int_{-\Theta/2}^{\Theta/2} \frac{\sin(\Omega(t-\tau))}{\pi(t-\tau)} \alpha(\omega, t) dt = e^{-i\omega\tau}, \quad \text{for } -\Theta/2 \leq \tau \leq \Theta/2.$$

## Derivation

### 1. Mathematical Context and Weak Formulation

Direct minimization of an $L^2$ error functional containing a Dirac delta distribution is not strictly defined because $\delta \notin L^2$ and its square $|\delta|^2$ diverges. To resolve this, we interpret the relation between the target distribution and the integral estimator in the weak sense.

Let the continuous-time synthesis function be defined as:

$$I(\omega_0) = \int_{-\Theta/2}^{\Theta/2} e^{i\omega_0 t} \alpha(\omega, t) dt.$$

We demand that $I(\omega_0)$ matches the target distribution $2\pi\delta(\omega - \omega_0)$ over the interval $\omega_0 \in [-\Omega, \Omega]$ when tested against an appropriate set of test functions. Choosing the complex exponentials $\varphi(\omega_0) = e^{-i\omega_0\tau}$ as our test functions (where $\tau \in [-\Theta/2, \Theta/2]$ acts as the coordinate index), the weak projection identity requires:

$$\int_{-\Omega}^{\Omega} I(\omega_0) e^{-i\omega_0\tau} d\omega_0 = \int_{-\Omega}^{\Omega} 2\pi\delta(\omega - \omega_0) e^{-i\omega_0\tau} d\omega_0, \quad -\Theta/2 \leq \tau \leq \Theta/2.$$

### 2. Evaluation of the Target Distribution Term

Applying the sifting property of the Dirac delta distribution to the right-hand side of the weak formulation yields:

$$\int_{-\Omega}^{\Omega} 2\pi\delta(\omega - \omega_0) e^{-i\omega_0\tau} d\omega_0 = 2\pi e^{-i\omega\tau},$$

provided that the parameter $\omega$ lies within the optimization band, i.e., interval $\omega \in [-\Omega, \Omega]$.

### 3. Evaluation of the Synthesis Term

Substituting the definition of $I(\omega_0)$ into the left-hand side of the weak formulation gives:

$$\int_{-\Omega}^{\Omega} \left( \int_{-\Theta/2}^{\Theta/2} e^{i\omega_0 t} \alpha(\omega, t) dt \right) e^{-i\omega_0\tau} d\omega_0.$$

Since the integration domains are bounded and the integrand is continuous with respect to all variables, Fubini's theorem permits the exchange of the order of integration:

$$\int_{-\Theta/2}^{\Theta/2} \alpha(\omega, t) \left( \int_{-\Omega}^{\Omega} e^{i\omega_0(t-\tau)} d\omega_0 \right) dt.$$

### 4. Integration of the Kernel and Final Setup

The inner integral over $\omega_0$ evaluates to a standard Dirichlet kernel:

$$\int_{-\Omega}^{\Omega} e^{-i\omega_0(t-\tau)} d\omega_0 = \left[\frac{e^{-i\omega_0(t-\tau)}}{i(t-\tau)}\right]_{-\Omega}^{\Omega} = \frac{e^{i\Omega(t-\tau)} - e^{-i\Omega(t-\tau)}}{i(t-\tau)} = \frac{2\sin\big(\Omega(t-\tau)\big)}{t-\tau}.$$

Substituting this result back into the rearranged integral yields:

$$\int_{-\Theta/2}^{\Theta/2} \alpha(\omega,t) \frac{2\sin\big(\Omega(t-\tau)\big)}{t-\tau} dt.$$

**5. Equating Both Sides**

Equating the evaluated left-hand and right-hand sides results in:

$$\int_{-\Theta/2}^{\Theta/2} \alpha(\omega,t) \frac{2\sin\big(\Omega(t-\tau)\big)}{t-\tau} dt = 2\pi e^{-i\omega\tau}.$$

Dividing both sides by $2\pi$ produces the final Fredholm integral equation of the first kind with a *Sinc* kernel:

$$\int_{-\Theta/2}^{\Theta/2} \frac{\sin\big(\Omega(t-\tau)\big)}{\pi(t-\tau)} \alpha(\omega,t) dt = e^{-i\omega\tau}, \quad -\Theta/2 \le \tau \le \Theta/2.$$

This completes the proof.

## APPENDIX B – Proof of (17.2) and (17.3)

This appendix provides a rigorous derivation showing that enforcing the target relation containing the Dirac delta distribution $\delta(\omega - \omega_0)$ in the weak sense (or via projections onto the Fourier basis) on the discrete least-squares functionals (14.2) and (14.3),

$$\Delta = \int_{-\Omega}^{\Omega} \left| 2\pi\delta(\omega-\omega_0) - \sum_{k=0}^{K-1} e^{i\omega_0 t_k} \alpha(\omega, t_k) \right|^2 d\omega_0,$$

$$\Delta = \int_{-\Omega}^{\Omega} \left| 2\pi\delta(\omega-\omega_0) - \sum_{k=0}^{K-1} e^{i\omega_0 kT} \alpha(\omega, kT) \right|^2 d\omega_0,$$

produces the system of linear equations (17.2) and (17.3) with a discrete *Sinc* kernel matrix:

$$\sum_{k=0}^{K-1} \frac{\sin(\Omega(t_k - t_l))}{\pi(t_k - t_l)} \alpha(\omega, t_k) = e^{-i\omega t_l},$$

$$\sum_{k=0}^{K-1} \frac{\sin(\Omega(k-l)T)}{\pi(k-l)T} \alpha(\omega, kT) = e^{-i\omega lT}.$$

## Derivation

**1. Mathematical Context and Weak Formulation**

As established in **Appendix A**, direct minimization of an $L^2$ error functional containing a Dirac delta distribution is not strictly defined because the square $|\delta|^2$ diverges. Instead, we project the relation between the target distribution and the discrete-time synthesis estimator onto the appropriate test function space.

Let the discrete-time synthesis function for arbitrary (nonuniform) sampling times $t_k$ be defined as:

$$I(\omega_0) = \sum_{k=0}^{K-1} e^{i\omega_0 t_k} \alpha(\omega, t_k).$$

We demand that $I(\omega_0)$ matches the target distribution $2\pi\delta(\omega - \omega_0)$ over the interval $\omega_0 \in [-\Omega, \Omega]$ in the week sense. For a discrete system with $K$ degrees of freedom, we choose the complex exponentials matching the sampling grid, $\varphi_l(\omega_0) = e^{-i\omega_0 t_l}$, for $l$ = 0,1, …, $K$-1, as our test functions. The weak projection identity requires that for each index $l$:

$$\int_{-\Omega}^{\Omega} I(\omega_0) e^{-i\omega_0 t_l} d\omega_0 = \int_{-\Omega}^{\Omega} 2\pi\delta(\omega - \omega_0) e^{-i\omega_0 t_l} d\omega_0, \quad l = 0,1, \dots, K-1.$$

**2. Evaluation of the Target Distribution Term**

Applying the sifting property of the Dirac delta distribution to the right-hand side yields:

$$\int_{-\Omega}^{\Omega} 2\pi\delta(\omega - \omega_0) e^{-i\omega_0 t_l} d\omega_0 \ = 2\pi e^{-i\omega t_l},$$

assuming that the frequency parameter $\omega$ falls within the optimization band $[-\Omega, \Omega]$.

**3. Evaluation of the Discrete Synthesis Term**

Substituting the definition of $I(\omega_0)$ into the left-hand side gives:

$$\int_{-\Omega}^{\Omega} \left( \sum_{k=0}^{K-1} e^{i\omega_0 t_k} \alpha(\omega, t_k) \right) e^{-i\omega_0 t_l} d\omega_0.$$

By the linearity of integration, we exchange the order of summation and integration to isolate the inner integral:

$$\sum_{k=0}^{K-1} \alpha(\omega, t_k) \left( \int_{-\Omega}^{\Omega} e^{i\omega_0 (t_k - t_l)} d\omega_0 \right).$$

**4. Integration of the Kernel and Matrix Formulation**

The inner integral over $\omega_0$ evaluates to a standard Dirichlet kernel:

$$\int_{-\Omega}^{\Omega} e^{-i\omega_0 (t_k - t_l)} d\omega_0 = \frac{2\sin\big(\Omega(t_k - t_l)\big)}{t_k - t_l}.$$

where for $k = l$, the expression is understood in the limit sense as $2\Omega$. Substituting this back into the summation yields:

$$\sum_{k=0}^{K-1} \alpha(\omega, t_k) \frac{2\sin\big(\Omega(t_k - t_l)\big)}{t_k - t_l}.$$

**5. Nonuniform Sampling Case (17.2)**

Equating the evaluated left-hand and right-hand sides results in:

$$\sum_{k=0}^{K-1} \alpha(\omega, t_k) \frac{2\sin\big(\Omega(t_k - t_l)\big)}{t_k - t_l} \ = \ 2\pi e^{-i\omega t_l}.$$

Dividing both sides by $2\pi$ produces the system of linear equations for the nonuniform sampling case, matching exactly equation (17.2):

$$\sum_{k=0}^{K-1} \frac{\sin(\Omega(t_k - t_l))}{\pi(t_k - t_l)} \alpha(\omega, t_k) = e^{-i\omega t_l}, \quad l = 0,1, \dots, K-1.$$

**6. Uniform Sampling Case (17.3)**

For the uniform sampling grid, we substitute $t_k = kT$ and $t_l = lT$, where $T$ is sampling period. The time difference becomes $t_k - t_l = (k - l)T$. Substituting this relation into (17.2) gives:

$$\sum_{k=0}^{K-1} \frac{\sin(\Omega(k-l)T)}{\pi(k-l)T} \alpha(\omega, kT) = e^{-i\omega lT}, \quad l = 0,1,\ldots,K-1.$$

This matches exactly the uniform discrete-time system equation (17.3), completing the proof.

# APPENDIX C – Proof of (18.1) and (18.2)

This appendix provides a rigorous derivation showing that enforcing the weighted target relation containing the Dirac delta distribution $2\pi\delta(\omega - \omega_0)$ and a spectral weight $S(\omega_0)$ in the weak sense on the discrete least-squares functionals (13.2) and (13.3),

$$\Delta = \int_{-\Omega}^{\Omega} \left| 2\pi S(\omega_0)\delta(\omega-\omega_0) - \sum_{k=0}^{K-1} S(\omega_0) e^{i\omega_0 t_k} \alpha(\omega, t_k) \right|^2 d\omega_0,$$

$$\Delta = \int_{-\Omega}^{\Omega} \left| 2\pi S(\omega_0)\delta(\omega-\omega_0) - \sum_{k=0}^{K-1} S(\omega_0) e^{i\omega_0 kT} \alpha(\omega, kT) \right|^2 d\omega_0,$$

leads to the linear systems (18.1) and (18.2):

$$\sum_{k=0}^{K-1} \left( \frac{1}{2\pi} \int_{-\Omega}^{\Omega} |S(\omega_0)|^2 e^{i\omega_0(t_k - t_l)} d\omega_0 \right) \alpha(\omega, t_k) = |S(\omega)|^2 e^{-i\omega t_l},$$

$$\sum_{k=0}^{K-1} \left( \frac{1}{2\pi} \int_{-\Omega}^{\Omega} |S(\omega_0)|^2 e^{i\omega_0(k-l)T} d\omega_0 \right) \alpha(\omega, kT) = |S(\omega)|^2 e^{-i\omega lT}.$$

## Derivation

**1. Mathematical Context and Weak Formulation**

As established in the preceding appendices, direct minimization of an $L^2$ error functional containing a Dirac delta distribution is not strictly defined because the square $|\delta|^2$ diverges. Instead, we project the relation between the weighted target distribution and the weighted discrete-time synthesis estimator onto an appropriate test function space.

Let the discrete-time synthesis function for arbitrary (nonuniform) sampling times $t_k$ be defined as:

$$I_S(\omega_0) = S(\omega_0) \sum_{k=0}^{K-1} e^{i\omega_0 t_k} \alpha(\omega, t_k).$$

We demand that $I_S(\omega_0)$ matches the weighted target distribution $2\pi S(\omega_0)\delta(\omega - \omega_0)$ over the interval $\omega_0 \in [-\Omega, \Omega]$ in the week sense. To account for the spectral weighting $S(\omega_0)$ we choose a weighted set of complex exponentials matching the sampling grid, $\varphi_l(\omega_0) = S^*(\omega_0)e^{-i\omega_0 t_l}$, for $l = 0,1, \ldots, K$-1, as our test functions. The weak projection identity requires that for each index $l$:

$$\int_{-\Omega}^{\Omega} I(\omega_0) S^*(\omega_0) e^{-i\omega_0 t_l} d\omega_0 = \int_{-\Omega}^{\Omega} 2\pi S(\omega_0)\delta(\omega-\omega_0) S^*(\omega_0) e^{-i\omega_0 t_l} d\omega_0, \quad l = 0,1,\ldots,K-1.$$

**2. Evaluation of the Weighted Target Distribution Term**

Applying the sifting property of the Dirac delta distribution to the right-hand side yields:

$$\int_{-\Omega}^{\Omega} 2\pi S(\omega_0)S^*(\omega_0)\delta(\omega-\omega_0)e^{-i\omega_0 t_l}d\omega_0 = 2\pi S(\omega)S^*(\omega)e^{-i\omega t_l} = 2\pi|S(\omega)|^2 e^{-i\omega t_l},$$

assuming that the frequency parameter $\omega$ falls within the optimization band $[-\Omega, \Omega]$.

**3. Evaluation of the Weighted Discrete Synthesis Term**

Substituting the definition of $I_S(\omega_0)$ into the left-hand side gives:

$$\int_{-\Omega}^{\Omega}\left(S(\omega_0)\sum_{k=0}^{K-1} e^{i\omega_0 t_k}\alpha(\omega,t_k)\right)S^*(\omega_0)e^{-i\omega_0 t_l}d\omega_0.$$

Combining the spectral weight terms $S(\omega_0)S^*(\omega_0) = |S(\omega_0)|^2$. By the linearity of integration, we exchange the order of summation and integration to isolate the inner integral:

$$\sum_{k=0}^{K-1}\alpha(\omega,t_k)\left(\int_{-\Omega}^{\Omega}|S(\omega_0)|^2 e^{i\omega_0(t_k-t_l)}d\omega_0\right).$$

**4. Nonuniform Sampling Case (18.1)**

Equating the evaluated left-hand and right-hand sides results in:

$$\sum_{k=0}^{K-1}\alpha(\omega,t_k)\left(\int_{-\Omega}^{\Omega}|S(\omega_0)|^2 e^{i\omega_0(t_k-t_l)}d\omega_0\right) = 2\pi|S(\omega)|^2 e^{-i\omega t_l}.$$

Dividing both sides by $2\pi$ produces the system of linear equations for the nonuniform sampling case, matching exactly equation (18.1):

$$\sum_{k=0}^{K-1}\left(\frac{1}{2\pi}\int_{-\Omega}^{\Omega}|S(\omega_0)|^2 e^{i\omega_0(t_k-t_l)}d\omega_0\right)\alpha(\omega,t_k) = |S(\omega)|^2 e^{-i\omega t_l}, \quad l = 0,1,\dots,K-1.$$

**5. Uniform Sampling Case (18.2)**

For the uniform sampling grid, we substitute $t_k = kT$ and $t_l = lT$, where $T$ is sampling period. The time difference in the exponential kernel simplifies to $t_k - t_l = (k-l)T$. Substituting this relation into (18.1) yields:

$$\sum_{k=0}^{K-1}\left(\frac{1}{2\pi}\int_{-\Omega}^{\Omega}|S(\omega_0)|^2 e^{i\omega_0(k-l)T}d\omega_0\right)\alpha(\omega,kT) = |S(\omega)|^2 e^{-i\omega lT}, \quad l = 0,1,\dots,K-1.$$

This matches exactly the uniform discrete-time system equation (18.2), completing the proof.

**6. Note on $S(\omega_0)$ Restrictions**

The spectral model $S(\omega_0)$ is strictly subject only to the requirement that its total energy within the optimization band $[-\Omega, \Omega]$ is bounded and non-zero everywhere (i.e., it possesses a strictly positive $L^2$ norm):

$$0 < \int_{-\Omega}^{\Omega}|S(\omega_0)|^2 d\omega_0 < \infty.$$

# APPENDIX D – EDFT MATLAB Code

```
function [F,S,f,Stopit,A]=edft(X,N,tk,I,W)

%EDFT  Extended Discrete Fourier Transform.
```

```
%
% EDFT computes an N–point Extended Discrete Fourier Transform (EDFT)
% of signal X, optionally nonuniformly sampled at times tk, and returns:
%     F  - N–point Fourier transform of X (complex spectrum)
%     S  - Amplitude spectrum estimate (complex spectrum)
%     f  - Frequency query points used in the transform
%     Stopit - Iteration status information (see below)
%     A  - Extended Fourier basis matrix such that F = X*A
%
% X may contain NaN entries, which are interpreted as missing samples
% and ignored during the weighted least-squares estimation.
%
% ---------------------------------------------------------------------
% SYNTAX
%
% F = edft(X)
%     If X contains NaN, EDFT iterates automatically. Otherwise, FFT is
%     used and F = fft(X).
%
% F = edft(X,N)
%     Runs EDFT for N–point output. If N is a vector, it is interpreted
%     as the frequency grid fn, and N = length(fn). If length(X) > N,
%     X is truncated and FFT(X,N) is returned when no NaN exist.
%
% F = edft(X,N,tk)
%     Computes EDFT for data sampled at times tk. If tk = [], uniform
%     sampling tk = 0:K-1 is assumed. If N is scalar, frequency points are:
%         fn = ifftshift(-ceil((N-1)/2):floor((N-1)/2))/N;
%
% F = edft(X,N,tk,I)
%     Limits the maximum number of iterations to I. Default I is set by
%     the global parameter 'Miteration'.
%
% F = edft(X,N,tk,I,W)
%     Initializes EDFT weighting with vector W (W proportional to S).
%     Default: W = ones(size(F)). W must contain ≥ length(X) nonzero values.
%
% [F,S] = edft(...)
%     Returns amplitude spectrum S. For FFT cases, S = F / length(X).
%
% [F,S,f] = edft(...)
%     Also returns frequency points f (Hz or normalized units).
%
% [F,S,f,Stopit] = edft(...)
%     Returns iteration status:
%       Stopit(1,:) - Iterations completed
%       Stopit(2,:) - Stop reason:
%                     0 : max iterations reached
%                     1 : resolution deviation > EPS_R / no solution
%                     2 : convergence tolerance < EPS_C (converged)
%       Stopit(3,:) - Algorithm used:
%                     0 : FFT special case
%                     1 : Uniform fast EDFT
%                     2 : Uniform EDFT with NaN
%                     3 : General uniform / nonuniform EDFT
%
% [F,S,f,Stopit,A] = edft(...)
%     Also returns the Extended Fourier basis matrix A.
%     For vector-row X: F = X*A;
%     For vector-column X: F = A*X;
```

```
%      For matrix X: F(:,l) = A(:,:,l) * X(:,l);
%      Before multiplying by A, NaN in X must be replaced with zeros.
%
%   No output arguments: EDFT plot of |F| and |S| versus frequency.
%
%   For matrix X, each column is processed independently. If N or tk
%   are vectors, they are applied consistently to all columns of X.
%
%   ---------------------------------------------------------------------
%   ALGORITHM
%
%   Inputs:
%      X  - Data samples, length K (may include NaN)
%      N  - Number of frequency points or frequency vector fn
%      tk - Sample times (uniform or nonuniform)
%      I  - Maximum iterations
%      W  - Iterative weighting (power estimate)
%
%   Fourier basis matrix:
%      E = exp( -1i * 2*pi * (tk.' * fn) );
%
%   Core update equations (per iteration):
%
%      R = E * diag(W / N) * E';      % Discrete correlation matrix
%      A = (R \ E) .* W;              % Extended Fourier basis
%      F = X * A;                     % N–point EDFT
%      S = (X * (R \ E)) ./ diag(E' * (R \ E)).'; % Amplitude spectrum
%      W = S .* conj(S);              % Iterate weight from power
%
%   Special case: if length(X) == N, EDFT is non-iterative and W-independent.
%
%   Fast algorithms (uniform sampling) use FFT/IFFT and Levinson–Durbin
%   recursion to accelerate inversion of R.
%
%   ---------------------------------------------------------------------
%   FEATURES
%
%   1) F is an N–point Fourier transform; PSD estimate:
%         PSD = abs(F).^2 / (N*T)
%
%   2) S provides amplitude and phase estimates of sinusoidal components.
%
%   3) Inverse EDFT extrapolates/interpolates X to full length N:
%         Y = ifft(F);  (if frequencies on FFT grid)
%         Y = iedft(F,f,tn); (general case, tn - sample times of length N)
%
%   4) Handles arbitrary frequency sets f (uniform or nonuniform).
%
%   5) Achieves frequency resolution improvement up to factor N/K:
%         0 < F./S <= N   and   sum(F./S) = N*K   (all iterations)
%
%   6) Supports missing samples: NaN in X indicate unavailable data.
%
%   ---------------------------------------------------------------------
%   INPUT SELECTION GUIDELINES
%
%   Sampling tk examples (T - sampling period, default T = 1 s):
%     • Uniform sampling: tk = (0:K-1)*T;
%     • Missing samples: NaN in X
%     • Jittered sampling: tk = ((0:K-1) + jitter*rand(1,K))*T;
```

```
%    • Additive sampling: t0=0; tk = tk-1 + (1 + jitter*rand)*T;
%    • Gapped data
%    • Level-crossing or adaptive sampling
%
%   Frequency grid guidelines (Fs - sampling frequency, default Fs = 1 Hz):
%    • fn = Fs*ifftshift(-ceil((N-1)/2):floor((N-1)/2))/N;
%    • User may arbitrarily define fn
%    • Must include full expected signal support
%    • Real signals require symmetric ± frequencies for real output
%
%   --------------------------------------------------------------------
%   See also: FFT, IFFT, IEDFT, NUFFT
%

% AUTHOR: Vilnis Liepins (vilnislp@gmail.com)
%
% RFERENCES (available on researchgate.net):
%
% [1] Vilnis Liepins, A method of spectrum evaluation applicable to
%   analysis of periodically and non regularly digitized signals.
%   Automatic Control and Computer Sciences, Vol.27, No.6, pp.46-52, 1993.
%
% [2] Vilnis Liepins, A spectral estimation method of nonuniformly sampled
%   band-limited signals. Automatic Control and Computer Sciences,
%   Vol.28, No.2, pp.52-58, 1994.
%
% [3] Vilnis Liepins, An algorithm for evaluation a discrete Fourier
%   transform for incomplete data. Automatic Control and Computer Sciences,
%   Vol.30, No.3, pp.20-29, 1996.
%
% NOTE: The first version of file (gdft.m) was submitted to fileexchange
%   on October 7, 1997 as Matlab 4.1 code.

% ====================== Set parameters for Stopit ======================
Miteration = 30;       % Limit for maximum number of iteration (Stopit 0)
EPS_R = [0.0001 1];     % Threshold for resolution deviation (Stopit 1)
EPS_C = [0.0001 1];     % Threshold for convergence tolerance (Stopit 2)
% Threshold is applied if the second element is equal to 1.
% ==================== Setup input/output arguments ====================
if nargin==0||isempty(X), error('Not enough input arguments.'),end
% Input argument X
if sum(any(isinf(X))), error('Inf is not allowed in X.'),end
if size(X,1)==1, X=X.';trf=1;else,trf=0;end
[K,L]=size(X);
% Input argument N
fn=[];
if nargin<2||isempty(N)
   N=K;
elseif sum(any(isnan(N)))||sum(any(isinf(N)))
   error('NaN or Inf is not allowed in N.')
elseif isscalar(N), N=floor(abs(N));    % N is a scalar
elseif sum(any(isnan(X)))
   error('NaN is not allowed in X if N is a vector.')
else
   if size(N,1)==1, N=N(:);end        % N was vector row
   fn=real(N);[N, NL]=size(fn);       % Set frequencies fn
   if NL~=L&&NL~=1, error('Incorrect size of vector N (fn).')
   elseif NL==1&&L>1,fn=ones(N,L).*fn;end  % fn - 2 dim array
end
if N<K, X=X(1:N,:);K=N;end         % Truncate X if more than N points
```

```
% Input argument tk
if nargin>2&&~isempty(tk)
   if sum(any(isnan(X)))||sum(any(isnan(tk)))||sum(any(isinf(tk)))
      error('NaN or Inf is not allowed in X and tk.')
   elseif size(tk,1)==1,tk=tk(:);
   end
   [TK, TL] = size(tk);tk=real(tk);
   if TK>K, tk=tk(1:K,:);end      % Truncate tk if has more than X points
   if TK<K||(TL~=L&&TL~=1), error('Incorrect size of tk.')
   elseif TL==1&&L>1,tk=ones(K,L).*tk;end      % tk - 2 dim array
   if isempty(fn)
      fn=ones(N,L).*(ifftshift(-ceil((N-1)/2):floor((N-1)/2))/N).';
   end
elseif ~isempty(fn),tk=ones(K,L).*(0:K-1).';    % Use default tk
end
% Input argument I
if nargin<4||isempty(I), I=Miteration;  % Set default value for I
else
   I=floor(abs(I(1)));
   if isnan(I)||isinf(I), error('NaN and Inf is not allowed in I.'),end
end
% Input argument W
if nargin<5||isempty(W), W=ones(N,L);   % Set default values for W
elseif sum(any(isnan(W)))||sum(any(isinf(W)))
   error('NaN and Inf is not allowed in W.')
else
   if size(W,1)==1, W=W(:);end
   [WN, WL] = size(W);
   if WN~=N||(WL~=L&&WL~=1), error('Incorrect size of W.')
   elseif WL==1&&L>1,W=ones(N,L).*W;   % W - 2 dim array
   end
   W=W.*conj(W);WK=sum(W>eps(max(W)));
   for l=1:L
      if WK(l)<K && WK(l)>0, W(:,l)=W(:,l)+max(W(:,l))*EPS_R(1);
      elseif WK(l)==0, W(:,l)=ones(N,1);end
   end
end
if nargout==5,Alf=1;W_P=ones(size(W));a=zeros(1,K);a_p=a.';
   INVR=zeros(K);RA=INVR;RE_P=zeros(K,N);A=zeros(N,K,L);
else,Alf=0;
end
F=zeros(N,L);S=F;f=F;XR=zeros(K,1);RE=XR;ERE=zeros(N,1);SW=zeros(1,I);
Stopit=[I*ones(1,L);zeros(2,L);];s = warning;warning('off','all');
R=zeros(K);
% ================== Process one by one each column of X ==================
for l=1:L
   Xnan=~isnan(X(:,l));   % Xnan indicate samples as '1' , NaN as '0'
   KK=sum(Xnan);          % KK the length of input data X w/o NaN
   if (~any(X(:,l))&&KK==K)||(KK==K&&K==N&&isempty(fn))||(KK==1&&K==1)||KK==0
      Alg=0;             % A special case - use FFT
      F(:,l)=fft(X(:,l),N);
      S(:,l)=F(:,l)/K;
      Stopit(:,l)=[1; 0; Alg;];
   elseif isempty(fn)&&KK==K
      Alg=1;             % Algorithm without NaN in uniform sequence
      Stopit(3,l)=Alg;
      for it=1:I
         Alg1_ERE;
         if EPS_R(2)==1 && stopit1,break,end    % Break if EPS_R reached
         F(:,l)=fft(XR,N);       % Calculate EDFT output
```

```
            calc_edft_out;
            if EPS_C(2)==1 && stopit2,break,end    % Break if EPS_C reached
        end
    elseif KK<K
        Alg=2;             % Algorithm with NaN in uniform sequence
        Stopit(3,l)=Alg;
        X(~Xnan,l)=zeros(K-KK,1);   % Replace NaN by 0 in X
        t=find(Xnan);              % Sample number vector
        INVR=zeros(K);
        for it=1:I
            Alg2_ERE;
            if EPS_R(2)==1 && stopit1,break,end    % Break if EPS_R reached
            F(:,l)=fft(conj(INVR)*X(:,l),N); % Calculate EDFT output
            calc_edft_out;
            if EPS_C(2)==1 && stopit2,break,end    % Break if EPS_C reached
        end
    else
        Alg=3;  % Algorithm for nonuniform/uniform sequence/frequency set
        if K==N,I=1;W=ones(N,l);end
        Stopit(:,l)=[I; 0; Alg;];
        E=exp(-1i*2*pi*tk(:,l)*fn(:,l).');  % Complex exponents matrix E
        for it=1:I
            Alg3_ERE;
            if EPS_R(2)==1 && stopit1,break,end    % Break if EPS_R reached
            F(:,l)=X(:,l).'*RE;     % Calculate EDFT output
            calc_edft_out;
            if EPS_C(2)==1 && stopit2,break,end    % Break if EPS_C reached
        end
    end
    if Alg~=3,f(:,l)=ifftshift(-ceil((N-1)/2):floor((N-1)/2)).'/N;
    else,f(:,l)=fn(:,l);end
    if Alf==1&&I>=1
        if Alg~=3,A(:,:,l)=exp(-1i*2*pi*f(:,l)*(0:K-1));
            if Alg==1,A(:,:,l)=A(:,:,l)*Alg1_RA.*W_P(:,l);end
            if Alg==2,A(:,:,l)=A(:,:,l)*RA.*W_P(:,l);end
        else,A(:,:,l)=RE_P.*W_P(:,l);
        end
    end
end
Stopit=table(Stopit,'RowNames',{'Iteration #';'Break Reason';'Algorithm #'});
if nargout==0
    clf
    for l=1:L
    if Alg~=3
        fpl=fftshift(f(:,l));
        Fp=abs(fftshift(F(:,l)));
        Sp=abs(fftshift(S(:,l)));
    else
        [fpl,ind]=sort(f(:,l));
        Fp=abs(F(ind,l));
        Sp=abs(S(ind,l));
    end
% Plots Extended DFT in subplot221.
    subplot(211)
    plot(fpl,Fp)
    xlabel('Frequencies in f (ascending)')
    ylabel('abs(F)')
    title('Extended DFT')
    hold on
% Plots Power Spectrum in subplot222.
```

```
    subplot(212)
    plot(fpl,Sp)
    xlabel('Frequencies in f (ascending)')
    ylabel('abs(S)')
    title('Amplitude Spectrum')
    hold on
    end
    hold off
end
if trf==1,F=F.';S=S.';f=f.';if Alf==1,A=A.';end,end;warning(s);
% =========================== Nested functions ===========================
function st1=stopit1
    stit=abs(ERE.'*W(:,l)/N/KK-1);st1=false;
    if (stit>EPS_R(1)&&it~=1)||isnan(stit)
        Stopit(1:2,l)=[it-1; 1;];st1=true;
    end
end
function st2=stopit2
    SW(it)=sum(W(:,l));st2=false;
    if it>1
        thit=abs(SW(it-1)-SW(it))/SW(1);
        if thit<EPS_C(1)
            Stopit(1:2,l)=[it; 2;];st2=true;
        end
    end
end
function calc_edft_out
    S(:,l)=F(:,l)./ERE;
    F(:,l)=F(:,l).*W(:,l);
    if Alf==1,W_P(:,l)=W(:,l);a_p=a';RA=conj(INVR);RE_P=RE.';end
    W(:,l)=S(:,l).*conj(S(:,l));
end
function Alg1_ERE          % Reference article [1]
    r=ifft(W(:,l));
    [a,V]=levinson(r,K-1);
    a=a.';rc=a;
    XR=zeros(K,1);RE=zeros(K,1);
    for k=1:K/2
        k0=K-k+1;k1=2:K-2*k+1;k2=k+1:K-k;k3=k:K-k+1;
        RE(1)=RE(1)+2*rc(k);
        RE(k0-k+1)=RE(k0-k+1)+2*rc(k0);
        RE(k1)=RE(k1)+4*rc(k2);
        XR(k)=XR(k)+rc(k3)'*X(k3,l);
        XR(k0)=XR(k0)+(flipud(rc(k3))).'*X(k3,l);
        XR(k2)=XR(k2)+rc(k2)*X(k,l)+flipud(conj(rc(k2)))*X(k0,l);
        rc(k2)=rc(k2-1)+conj(a(k+1))*a(k2)-a(k0)*flipud(conj(a(k2+1)));
    end
    if mod(K,2)==1
        RE(1)=RE(1)+rc(k+1);XR(k+1)=XR(k+1)+X(k+1,l)*rc(k+1);
    end
    ERE=real(fft(RE,N));W(:,l)=W(:,l)/real(V);
end
function Alg2_ERE          % Reference article [3]
    RT=ifft(W(:,l));
    R=toeplitz(RT(1:K));
    INVR(t,t)=R(t,t)\eye(KK);
    RE(1)=trace(INVR);
    for k=1:K-1
        RE(k+1,1)=sum(diag(INVR,k)+conj(diag(INVR,-k)));
    end
```

```
    ERE=real(fft(RE,N));
end
function Alg3_ERE          % Reference article [2]
    R=E*diag(W(:,l)/N)*E';
    RE=R\E;
    ERE=sum(conj(E).*RE).';
end
function RA=Alg1_RA     % Reference article [1]
    RA=zeros(K);RA(1,:)=a_p;RA(:,1)=a_p';
    RA(:,K)=flip(a_p);RA(K,:)=flip(a_p');
    for j=1:ceil(K/2)
        for k=j:K-j-1
            RA(j+1,k+1)=RA(j,k)+conj(a_p(j+1))*a_p(k+1)-a_p(K-j+1)*conj(a_p(K-k+1));
            RA(k+1,j+1)=conj(RA(j+1,k+1));
            RA(K-k,K-j)=RA(j+1,k+1);
            RA(K-j,K-k)=conj(RA(K-k,K-j));
        end
    end
end
end
```

# APPENDIX E – IEDFT MATLAB Code

```
function [Y,t]=iedft(F,fn,tn)

%IEDFT  Inverse Extended Discrete Fourier Transform.
%
%   IEDFT evaluates the inverse Fourier transform of EDFT output F or FFT
%   at specified sample times tn, corresponding to frequency query points fn.
%   Both fn and tn may be uniform or nonuniform grids.
%
%   ----------------------------------------------------------------------
%   SYNTAX
%
%   Y = iedft(F)
%       Performs IFFT(F) when frequency grid matches FFT grid spacing.
%       Y has the same size as F.
%
%   Y = iedft(F,fn)
%       Performs inverse transform at frequencies fn. If fn = [] or a scalar
%       equal to N = length(F), then IFFT grid is assumed:
%           fn = ifftshift(-ceil((N-1)/2) : floor((N-1)/2)) / N;
%
%   Y = iedft(F,fn,tn)
%       Computes inverse EDFT at arbitrary sample times tn.
%       If tn = [], default is tn = 0:N-1 (uniform sampling).
%
%   [Y,t] = iedft(...)
%       Also outputs the sample-time vector actually used.
%
%   No output arguments: plots real and imaginary parts of Y versus t.
%
%   For matrix input F, each column is processed independently. If fn or tn
%   are vectors, they are applied consistently to all columns of F.
%
%   ----------------------------------------------------------------------
%   ALGORITHM
%
```

```
%   Inputs:
%       F  - N–point Fourier spectrum (complex), EDFT/FFT result
%       fn - Frequency query points (uniform or nonuniform)
%       tn - Output sample times for Y (uniform or nonuniform)
%
%   Fourier synthesis matrix:
%       E = exp( 1i * 2*pi * (tn.' * fn) );
%
%   Core inverse transform:
%       Y = (F * E) / N;
%
%   Fast execution:
%       • Uses IFFT when fn is scalar equal to N or [], otherwise
%       • Uses NUFFT
%
%   ---------------------------------------------------------------------
%   REMARKS & USAGE NOTES
%
%   • For real-valued results, fn must contain symmetric +/– frequency pairs
%   • IEDFT provides interpolation and extrapolation when applied to EDFT
%     output F
%   • Best used as exact inverse of EDFT output :
%         Y = iedft( edft(X,N) );              (uniform case)
%         Y = iedft( edft(X,fn,tk), fn, tn );   (nonuniform case)
%
%   ---------------------------------------------------------------------
%   See also: EDFT, FFT, IFFT, NUFFT
%

% AUTHOR: Vilnis Liepins (vilnislp@gmail.com)
%
% REFERENCE: Vilnis Liepins. Extended Fourier analysis of signals. 2013.

% ==================== Setup input/output arguments ====================
% Check input argument F
if nargin<1||isempty(F),error('Not enough input arguments.')
elseif sum(any(isnan(F)))||sum(any(isinf(F)))
    error('NaN and Inf is not allowed in F.')
end
if size(F,1)==1,trf=1;F=F.';else,trf=0;end
[N,L]=size(F);
% IFFT applied to F if fn, tn are not vectors
if nargin==1|| ...
    nargin==2&&(isempty(fn)||(isscalar(fn)&&fn==N))|| ...
    nargin==3&&isempty(tn)&&(isempty(fn)||(isscalar(fn)&&fn==N))
    Y=ifft(F);
    t=ones(N,L).*(0:N-1).';
    if trf==1,Y=Y.';t=t.';end          % Adjust size of output
    if nargout==0,plot_Y_t,end
    return
% Check input argument fn
elseif sum(any(isnan(fn)))||sum(any(isinf(fn)))
    error('NaN or Inf is not allowed in fn.')
elseif isempty(fn)||(isscalar(fn)&&fn==N)
    fn=ones(N,L).*(ifftshift(-ceil((N-1)/2):floor((N-1)/2))/N).';
else
    if size(fn,1)==1,fn=fn(:);end       % fn was vector row
    fn=real(fn);
    if size(fn,1)~=N,error('Incorrect size of fn.'),end
    if size(fn,2)==1&&L>1,fn=ones(N,L).*fn;end  % fn is 2 dim array
```

```
    if size(fn,2)~=L,error('Incorrect size of fn.'),end
end
% Check input argument tn
if nargin<3||isempty(tn)
    tn=ones(N,L).*(0:N-1).';
elseif sum(any(isnan(tn)))||sum(any(isinf(tn)))
    error('NaN or Inf is not allowed in tn.')
else
    if size(tn,1)==1, tn=tn(:);end      % tn was vector row
    tn=real(tn);[TR,TC]=size(tn);
    if TC~=L&&TC~=1,error('Incorrect size of tn.'),end
    if TC==1&&L>1,tn=ones(TR,L).*tn;end % tn is 2 dim array
end
Y=zeros(size(tn));                     % Set default values for Y
% ================== Process one by one each column of F ==================
for l=1:L
    Y(:,l)=nufft(F(:,l),fn(:,l),-tn(:,l))/N;
end
if trf==1,Y=Y.';t=tn.';else,t=tn;end   % Adjust size of output
if nargout==0,plot_Y_t,end
% =========================== Nested function ============================
function plot_Y_t
    clf
% Plots Real part of Y in subplot221.
    subplot(211)
    plot(t,real(Y))
    xlabel('Time (t)')
    ylabel('Magnitude')
    title('Real part of IEDFT')
% Plots Imaginary part of Y in subplot222.
    subplot(212)
    plot(t,imag(Y))
    xlabel('Time (t)')
    ylabel('Magnitude')
    title('Imaginary part of IEDFT')
end
end
```